# Unveiling the local elemental arrangements across the interfaces inside CdSe/Cd$_{1-x}$Zn$_x$S core-shell and CdSe/CdS/ Cd$_{1-x}$Zn$_x$S core-crown-shell quantum wells

*Tatiana Lastovina[1], Oleg Usoltsev[2], Furkan Işik[1], Andriy Budnyk[1], Messaoud Harfouche[3], Betul Canımkurbey[1], and Hilmi Volkan Demir[1, 4,\*]*

[1] Department of Electrical and Electronics Engineering, Department of Physics, UNAM − Institute of Materials Science and Nanotechnology and The National Nanotechnology Research Center, Bilkent University, Ankara 06800, Turkey.

[2] ALBA Synchrotron, Carrer de la Llum, 2, 26, 08290 Cerdanyola del Vallès, Barcelona, Spain.

[3] Synchrotron-light for Experimental Science and Applications in the Middle East (SESAME), P.O. Box 7, Allan 19252, Jordan.

[4] Luminous! Center of Excellence for Semiconductor Lighting and Displays, School of Electrical and Electronic Engineering, Division of Physics and Applied Physics, School of Physical and Mathematical Sciences, School of Materials Science and Engineering, Nanyang Technological University, Singapore 639798, Singapore.

ABSTRACT

Colloidal CdSe-based nanoplatelets (NPLs) have found various applications in optoelectronics thanks to their bright and narrow photoluminescent (PL) emission. The peak emission wavelength of their PL is conveniently tunable and controllable through structural parameters such as the number of monolayers (ML) in the core, framing the CdSe core with a wider bandgap CdS crown, and covering either core or core-crown with a gradient Cd$_{1-x}$Zn$_x$S shell. The resulting heterostructures contain the core-shell interfaces, which significantly affect the PL emission efficiency. The internal structure and related atomic distribution within the interface region of NPLs are hardly accessible with conventional lab-scale instruments. Such critical studies rely on X-ray absorption spectroscopy (XAS), possibly available at synchrotron radiation facilities. We report on a systematic study of the Cd, Zn, Se, and S elemental distributions across the interfaces in CdSe/Cd$_{1-x}$Zn$_x$S core-shell and CdSe/CdS/Cd$_{1--x}$Zn$_x$S core-crown-shell quantum wells (QWs) with the CdSe core thickness ranging from 3.5 to 5.5 ML. By processing the XAS data (X-ray absorption near edge structure (XANES) and extended X-ray absorption fine structure (EXAFS)), we observe that the Cd–Se bonds dominate at the CdSe/Cd$_{1-x}$Zn$_x$S core-shell interface of structures with the 3.5 ML cores, while the Cd–Se bonds were more abundant in the cases of the 4.5 and 5.5 ML cores. The complementary information about prevailing bonds were extracted for other constituting elements, thus, describing the distribution of the elements at the core-shell interface of CdSe-based NPLs. The naked CdSe cores are covered with an organic shell (mostly oleic acid) via bridging oxygen atoms. Also, we address the issue of stability of such core-shell systems over the time. We demonstrate that after a half year of aging of the commercial-ready 4.5 ML CdSe/Cd$_x$Zn$_{1-x}$S NPLs, the Cd–Se bonds become more evident due to the partial degradation of the Cd–S bonds. This is the first experimental assessment of prevailing interatomic bonds at the core-shell interface in the CdSe-based NPLs of incremental structural heterogeneity, providing factual evidences about the elemental arrangement inside the core-crown-shell NPLs and the growth path of crowns and shells. Such structural information is of paramount importance for better understanding the relationship between the structure and electronic (emissive) properties of NPLs.

**INTRODUCTION**

Colloidal nanocrystals (NCs) have been known from the time of Faraday's gold sols (1857), but the interest in semiconductor quantum dots (QDs) has been driven by Alexey Ekimov's studies on quantum size effect in the 1980s [1]. The overall history of NCs' employment in different crafts spans over millennia [2] but has been appreciated only recently thanks to our understanding of the properties of these nano-objects. The field has been growing rapidly thanks to many decisive contributions such as Louis Brus' theoretical works on the excited electronic states of semiconductor crystallites [3] and Moungi Bawendi's experimental works on the hot-injection synthesis of QDs [4]. Their importance has been apprised with the Nobel Prize in Chemistry in 2023 awarded to those three pioneers [5].

Nowadays the field is thought relatively mature, and II–VI metal chalcogenides can be synthesized with tailored shapes, elemental composition, and heterostructure design. Among different form-factors, colloidal CdSe nanoplatelets (NPLs), efficiently combining advantages of colloidal QDs and epitaxial quantum wells (QWs), constitute the narrowest room-temperature emitters that have been employed in light-emitting diodes (LEDs), lasers, solar cells, etc.[6] The growing demand for customized materials for new types of optoelectronic devices encourages an attentive inquiry into the impact every structural component has on the resulting optical properties of NPLs.

Our group has long been active in synthesizing metal chalcogenide NPLs by wet chemistry methods [7]. The practiced approaches are based on the high-temperature decomposition of organometallic precursors, while the shell/crown is grown by either the hot injection shell method [8] or by the colloidal atomic layer deposition (c-ALD) [9], the latter being conducted at ambient or moderate temperatures [10]. Colloidal NPLs are atomically flat, quasi-2D sheets with semiconducting properties. In a typical case of CdSe, the core can be framed or covered by a wider band gap CdS, ZnS, or CdZnS compounds, called a crown or a shell, respectively. The approach sets a plethora of stable and bright heterostructures, including the core/shell CdSe/CdS and CdSe/CdZnS NPLs[11], the core-crown CdSe/CdS NPLs [12], and the CdSe/CdS@CdS core-crown/shell NPLs [13]. The thickness of either core and shell (hence, the resulting PL) can be quantitatively tuned by selecting the appropriate number of monolayers (ML). The lateral shape of NPL can be square or rectangular; the large and thin NPLs may roll in tubes [14].

Even though synthesis techniques have been mastered well, the mechanism behind the formation of a highly anisotropic shape and reaching the precise atomic-scale thickness of NPLs remains speculative. It is ascribed to an intrinsic instability in growth kinetics that governs the formation of anisotropic shapes from isotropic crystal clusters (such as zincblende CdSe)[15]. Such opinion relies on calculations performed under the general theory of 2D nucleation and growth and matching experimental data from conventional optical spectroscopy, XRD, and TEM. It has been further solidified with *in situ* synchrotron-based small-angle and wide-angle X-ray scattering (SAXS/WAXS) probing of ultrathin CdSe NPLs, which revealed that the continuous lateral growth of NPLs has been fed by reactive monomers [16]. SAXS together with extended X-ray absorption fine structure (EXAFS) spectroscopy, yet another synchrotron-based method, has been successfully utilized to study chemical mechanisms of particle formation as summarized in a dedicated review by Whitehead and Finke [17].

Core/shell structuring introduces a passivation layer onto the core using shell materials with enhanced chemical stability. However, there is a lattice mismatch of CdSe with both CdS (4%) and ZnS (12%), causing some strain at the interface [18]. Interface defects stemming from the stress accumulation may contribute to the multiexciton recombination decreasing the optical gain lifetime [19]. To alleviate local elastic distortion more complex shell structures such as ternary core/shell/shell (e.g., CdSe/CdS/ZnS) and core/composition-gradient shell such as CdSe/CdS/$Cd_{1-x}Zn_xSe_{1-y}S_y$ were composed [20]. Such structural complexity makes elucidating the internal structure and related atomic distribution within the interface region an overarching challenge. It has been addressed by applying complementary analytical techniques of electron microscopy and atom probe tomography [21], combined experimental and theoretical analysis of Raman and infrared spectra [22], and more noticeably by joining the EXAFS and Raman spectroscopies to quantify the width of the material gradient at the core-shell interface [23].

To date there are no reported X-ray absorption spectroscopy (XAS) studies on a representative selection of CdSe-based NPLs with different core/shell architectures. The interfacial core/crown and core/shell regions are of particular importance in this regard due to their direct impact on luminescent properties. Scattered examples represent valuable contributions but focused rather on cores, like the observation of size-dependent structural disorder in CdSe nanocrystals by X-ray absorption near-edge spectroscopy (XANES) [24] or revealing the actual alloyed $Cd_xZn_{1-x}Te$ core in ZnTe/CdSe core/shell QDs by a combined XAS–TEM study [25]. The local atomic structure order, chemical composition, and nature of alloying in CdSe/ZnS QDs have been already studied at the XAFS beamline of the SESAME synchrotron [26], and this fact has contributed positively to setting our experiment.

In this study the synchrotron $K$-edge XAS measurements of Zn, Cd, and Se were used for a series of heterostructures with incremental complexity: CdSe core, CdSe/CdS core/crown, CdSe/$Cd_{1-x}Zn_xS$ core/gradient shell, and CdSe/CdS/$Cd_{1-x}Zn_xS$ core/crown/gradient shell. The cores were made of 3.5, 4.5, and 5.5 ML. We performed the combined analysis of collected spectra for both XANES and EXAFS regions to extract information about the oxidation state and the first-shell bond distances. Careful observations allowed us to elucidate the contributions from surface/interface atoms and obtain important insights about the material gradient in the shell.

## 2. RESULTS AND DISCUSSION

### 2.1. NPLs structure and morphology

For this systematic study, four sets of NPLs of different structures were prepared by a classical wet chemistry method in organic solvents under a protective atmosphere. The details are reported in SI and will be further recalled in the respective chapter when discussing the particular structure later. Also, the structural properties of similar NPLs have been already reported in several recent works from our group, in this section, we briefly discuss the dominant crystalline phase and the shape of NPLs.

CdSe can form cubic (zincblende, ZB, $F\bar{4}3m$) and hexagonal (wurtzite, WZ, $P6_3mc$) phases [27]. It is known that clusters and small nanoparticles adopt the cubic-like shape of the ZB phase, which with further growth evolves into other shapes changing the phase too. WZ is accounted as a more stable phase, and there are reported cases of phase transition from ZB to WZ over time link. The issue of stability of prepared NPLs will be also addressed here in the final section, dedicated to aging.

The XRD patterns were measured on films of drop-casted fresh samples after evaporation of a solvent (SI, Fig. S1-S2). Their profile is affected by a relatively low quantity of diffracting medium and by size effects, known for nanoparticles. The analysis of the peak positions suggests that CdSe NPLs crystallize into the ZB structure.

TEM images (Fig. S3) demonstrate rectangular-shaped NPLs, known for this type of materials [28] [29]. In the case of 3.5 ML CdSe and 3.5 ML CdSe/CdS NPLs, the particles appear rolled along the longer size or diagonally. Such belt-like appearance is expected for very thin NPLs and it is in a good agreement with how similar NPLs look based on the literature data [16, 30]. Due to their lightweight, these belts can move under the electron beam making their TEM observation and acquisition of the SAED diffraction patterns challenging tasks. However, the growth of the ZnS shell leads to unfolding of scrolled 3.5 ML CdSe and 3.5 ML CdSe/CdS NPLs. The mean sizes of NPLs were obtained by graphically processing TEM images (Table. S1). They are used to define the structure of NPLs for each core thickness (3.5-5.5 ML) and. For instance, for NPLs of the CdSe/CdZnS structure, the mean length values are 7.7, 16.1 and 41.1 nm for 3.5, 4.5 and 5.5 ML, respectively. Instead, for the CdSe/CdS/CdZnS structure this row appears as 18.3, 29.8, and 60.3 nm (accordingly for 3.5-5.5 ML).

Thus, the stoichiometry of Cd and Se in CdSe NCs still remains an open question because of challenging assessment with conventional experimental techniques. Thus, such stoichiometry has been carried inquired in rare cases and is poorly addressed in the literature. The PL spectra of the samples are reported in SI (Fig. S4) for the reference.

## 2.2. Probing the Composition and Local Atomic Arrangement by XAFS
### 2.2.1. CdSe core NPLs

The mechanism favoring an anisotropic growth of the CdSe NPLs remains under discussion. Obviously, the nature of precursors, and synthesis parameters (temperature, reaction time, reagents concentration, etc.) each have their impact on the morphology of the final products. However, it is unclear why the facets grow at a different rate forming the anisotropic NPLs. The group of Prof. Norris [31] proposed the following mechanism based on the *in situ* TGA/DTA analysis of the CdSe NPLs growth from $Cd(propionate)_2$ and Se solid mixtures: (i) the formation of bis(propionyl) selenide, a reactive intermediate, (ii) that further reacts with $Cd(propionate)_2$ to yield CdSe and propionate/propionyl ions; (iii) the propionyl cations later decompose to carbon monoxide, ethene, and protons; (iv) possibility of propionate to attract protons to form propionic acid. It should be noted that the formation of CdO was also detected. Instead, a melt synthesis demonstrates how NPLs form in a highly concentrated environment [32]. Adding a short-chain carboxylate can cause a phase separation of the cadmium precursors, leading to their high local concentration. The surface kinetics (the rate of monomer addition /removal) will determine the shape of growing particles.

The 4.5 ML and 5.5 ML CdSe NPLs were prepared based on the well-known protocol [33]. It starts with long-chained cadmium carboxylates and elemental Se dissolved in 1-octadecene (ODE). After nucleation of small CdSe seeds short-chained cadmium carboxylates have been added to promote lateral growth of 2D nanocrystals. We synthesized 4.5 ML and 5.5 ML CdSe NPLs using both Cd(myristate)$_2$ and Cd(OAc)$_2$ as the Cd precursors. It is worth outlining that Cd(myristate)$_2$ dissolves completely in ODE at 180 °C, whereas the 1:1 mixture of Cd(myristate)$_2$ and Cd(OAc)$_2$ do not. This fact suggests that Cd(OAc)$_2$ may be used to induce a phase separation from a solvent [15]. In the case of 3.5 ML CdSe, the well-shaped NPLs were obtained with the addition of a short-chained Cd(OAc)$_2$·2H$_2$O only. Also, Se was used in the form of Se solution in 1-ODE. The proposed mechanism includes the following steps: thermally initiated by the homolytic cleavage of the Se–Se bond, the attack of the allylic proton in 1-ODE by Se radicals leading to the migration of the double bond and subsequent H transfer from the Se intermediate species [34]. The final products are dialkyl polyselenides, according to the K-edge Se XAFS spectrum and FEFF calculations [35], remained stable during holding [36].

The surface of NPLs is covered with a loose organics shell consisting either of oleic acid or acetates or a mixture thereof bonded to the particle's surface via carboxylic groups [37]. This implies the presence of bridging oxygen next to the surface, which affects the state of surface metal atoms. Thus, in order to fit the Cd-edge EXAFS data, we had to take into account a contribution from a possible surface bonded oxygen species. XANES spectra of Cd *K*-edge demonstrate a gradual decrease of the white line intensity with the increase of the number of layers.

It was revealed that samples contain contributions from the Cd–Se and Cd–O shells. The contribution from the Cd–Se bonds has been remained constant through varying of layers. The Cd–O coordination number has been found to decrease with increasing the number of layers. This effect can be described by the decrease of a relative percentage of surface atoms with a dominating contribution from atoms in the core of NPLs. It should be noted that both Cd–Cd and Cd–Se bonds have indistinguishable contributions to the shape of the Fourier transform (FT) of EXAFS spectrum.

The analysis of the Se *K*-edge spectra allows one to conclude that the Cd–Se interface remains constant with the layer variation. Also, there is no evidence of Se coordinated with light atoms like O (with expected interatomic distance around 2.3 Å). The details of XAFS analysis are presented in Fig. S6. It is possible to conclude that all Se atoms are bonded with Cd, while part of Cd is bonded with lighter atoms. The later can be described by the fact that O coordination can be realized through Se or Cd surface atoms, which is supported by the published results for CdSe QDs [38] showing that the Cd/Se atomic ratio could be greater than 1. Due to passivating ligands more Cd than Se atoms are located on the surface. Also, advanced nuclear magnetic resonance and infrared spectroscopy techniques confirmed the passivation of the Cd surface atoms with carboxylate in the case of ZB CdSe NPLs [39] [40]. The answer to the question of NPLs' organic shell composition lies beyond the scope of this work. We will limit here with the assumption that they are covered with oleic acid as evidenced by FTIR measurements (Fig. S5). Since the carboxylic vibration from both acetates and oleic acid head group may match in the spectrum, a minor presence of acetates could not be fully excluded.

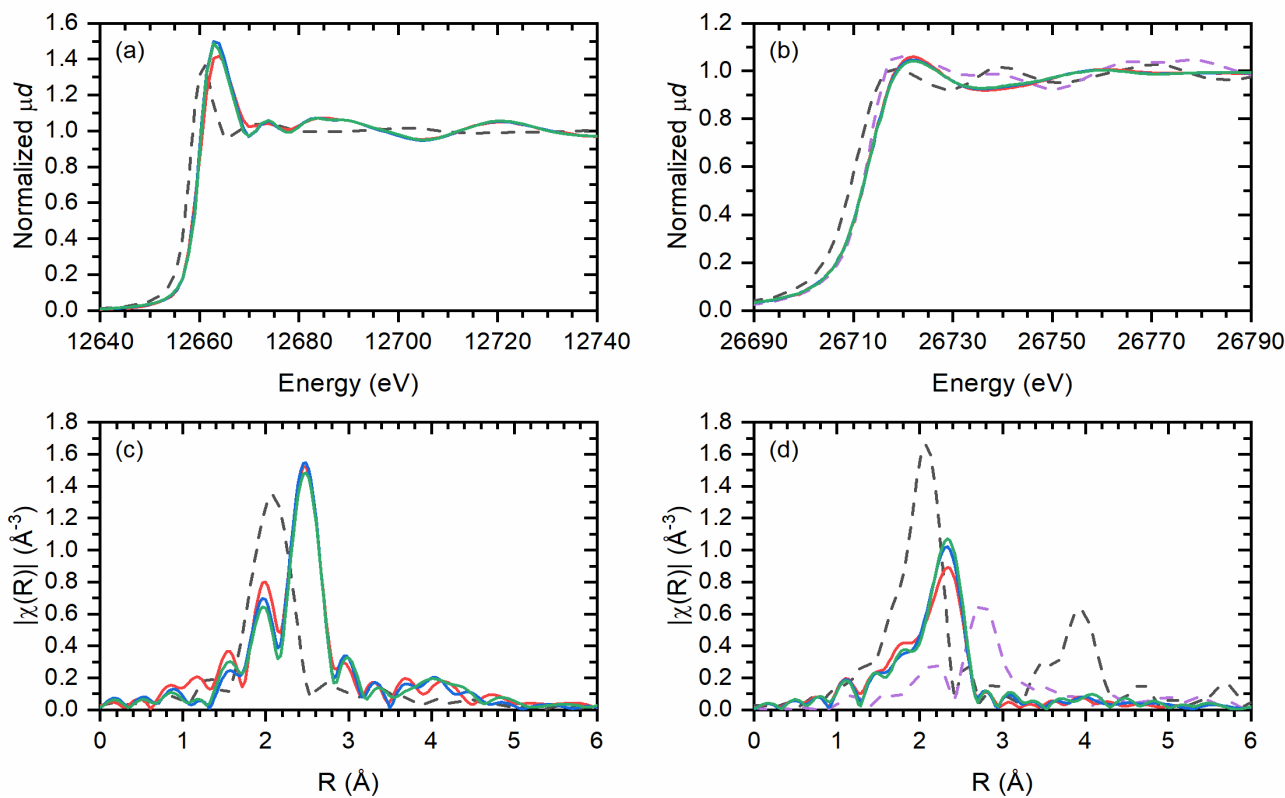

**Figure 2**. XANES spectra of CdSe NPLs with 3.5 (red lines), 4.5 (blue lines) and 5.5 ML (green lines) at Se $K$-edge (a) and Cd $K$-edge (b) and respective Fourier transformed of EXAFS (c and d).

**Table 1.** Parameters extracted from the standard 1$^{st}$ shell FT analysis of Se K-edge.

| Name | Bonding | R (Å) | R err. (Å) | N | N err. | $\sigma^2$ | $\sigma^2$ err. |
|---|---|---|---|---|---|---|---|
| 3.5 ML CdSe | Se-Cd | 2.6139 | 0.0043 | 3.126 | 0.241 | 0.0051 | 0.0006 |
| 4.5 ML CdSe | Se-Cd | 2.6168 | 0.0038 | 3.245 | 0.222 | 0.0053 | 0.0005 |
| 5.5 ML CdSe | Se-Cd | 2.6200 | 0.0048 | 3.028 | 0.263 | 0.0049 | 0.0006 |

**Table 2.** Parameters extracted from the standard 1$^{st}$ and 2$^{nd}$ shell FT analysis of Cd K-edge.

| Name | Bonding | R (Å) | R err. (Å) | N | N err. | $\sigma^2$ | $\sigma^2$ err. |
|---|---|---|---|---|---|---|---|
| 3.5 ML CdSe | Cd-Se | 2.6204 | 0.0063 | 3.165 | 0.335 | 0.0068 | 0.0007 |
|  | Cd-O | 2.2480 | 0.0402 | 2.252 | 0.731 | 0.0183 | 0.0075 |
| 4.5 ML CdSe | Cd-Se | 2.6218 | 0.0047 | 3.416 | 0.266 | 0.0066 | 0.0005 |
|  | Cd-O | 2.2305 | 0.0531 | 2.172 | 0.979 | 0.0237 | 0.0105 |
| 5.5 ML CdSe | Cd-Se | 2.6273 | 0.0044 | 3.303 | 0.269 | 0.0062 | 0.0005 |
|  | Cd-O | 2.1990 | 0.0456 | 1.341 | 0.531 | 0.0150 | 0.0081 |

**2.2.2. CdSe/Cd$_x$Zn$_{1-x}$S NPLs**

The CdSe/Cd$_x$Zn$_{1-x}$S NPLs samples represents CdSe core layers variation similar to one presented in previous chapter, but with Cd$_x$Zn$_{1-x}$S coating. The hot-injection approach for the shell growth includes the following steps: washing of the CdSe cores with ethanol, re-dispersing of the n ML CdSe (n=3.5, 4.5, 5.5) cores in 1-ODE with Cd and Zn precursors, degassing followed by addition of OAm to slow down the dissolution of CdSe core, bringing the reaction mixture to 300 °C while starting injection of the shell precursor [41] after 162 °C has been reached. The gradient shell formation is important because PL QY may reach the value of 95-98%.

Figure 3 shows XANES spectra of CdSe/Cd$_x$Zn$_{1-x}$S NPLs with 3.5, 4.5, and 5.5 ML CdSe cores at Cd *K*-edge, Se *K*-edge, and Zn *K*-edge and respective Fourier transformed curves of EXAFS. The results of the EXAFS data analysis (Fig. S8) for the Cd *K*-edge demonstrated the Cd-Se and Cd-S bonds presented for all ML CdSe/Cd$_x$Zn$_{1-x}$S NPLs samples. In the case of 3.5 ML CdSe/Cd$_x$Zn$_{1-x}$S NPLs Cd-Se bonds are dominant (N = 4.68) with a minor contribution of Cd coordinated to S (N = 1.66), while 4.5 and 5.5 ML CdSe/Cd$_x$Zn$_{1-x}$S NPLs samples contain mainly the Cd-S bonds (Table 3). The Cd/Zn interface is important since for the 3.5 ML CdSe/Cd$_x$Zn$_{1-x}$S sample surface layers of Zn are presented by ZnO, while 4.5 and 5.5 ML samples have Zn atoms coordinated with S (Fig. 4 c, f and Table 5). This might be observed by positions for Zn *K*-edge XANES spectra of 4.5 and 5.5 ML CdSe/Cd$_x$Zn$_{1-x}$S NPLs that are identical, while 3.5 ML one is shifted towards higher energy and has a different shape of spectrum features. The samples of 4.5 and 5.5 ML CdSe/Cd$_x$Zn$_{1-x}$S NPLs has features similar to those of the ZB cubic lattice [26], while 3.5 ML CdSe/Cd$_x$Zn$_{1-x}$S NPLs contain dominant oxygen bonding. In all three samples, the edge remains at the same energy position indicating that Zn is in its formal 2+ oxidation state. The difference between Zn-S and Zn-O supported by FT intensities of the k$^2$-weighted Zn *K*-edge EXAFS (Fig. 4f) for the *k*-range of 3 to 10 Å$^{-1}$. The main peak near R = 1.9 Å (phase uncorrected) in the case of 4.5 and 5.5 ML CdSe/Cd$_x$Zn$_{1-x}$S NPLs corresponds to the Zn-S bonds and a 4-coordination of sulfur atoms similar to ZnS crystal lattice, whereas small peaks near R = 1.3 Å (phase uncorrected) correspond to the Zn-O bonds. Also, the peak at around 3.2 Å (phase uncorrected) corresponds to the 2$^{nd}$ shell around the Zn atoms that were previously observed [42, 43], but cannot be properly identified because of the quality of the signal with respect to the other samples. The Zn-S bond lengths for the 4.5 and 5.5 ML CdSe/Cd$_x$Zn$_{1-x}$S NPLs are 2.34 and 2.36, which is close to the bulk ZnS ZB phase (2.35 Å). It is possible when the ZnS shell does not experience a strong impact from the core. It should be noted that 5.5 ML CdSe/Cd$_x$Zn$_{1-x}$S has more mixing in the narrow bandgap CdSe core with S from ZnS and the formation of CdSeS interface. Such arrangement may be responsible for a blue shift in their PL emission (Fig. S4), similar to what was observed by other groups [26].

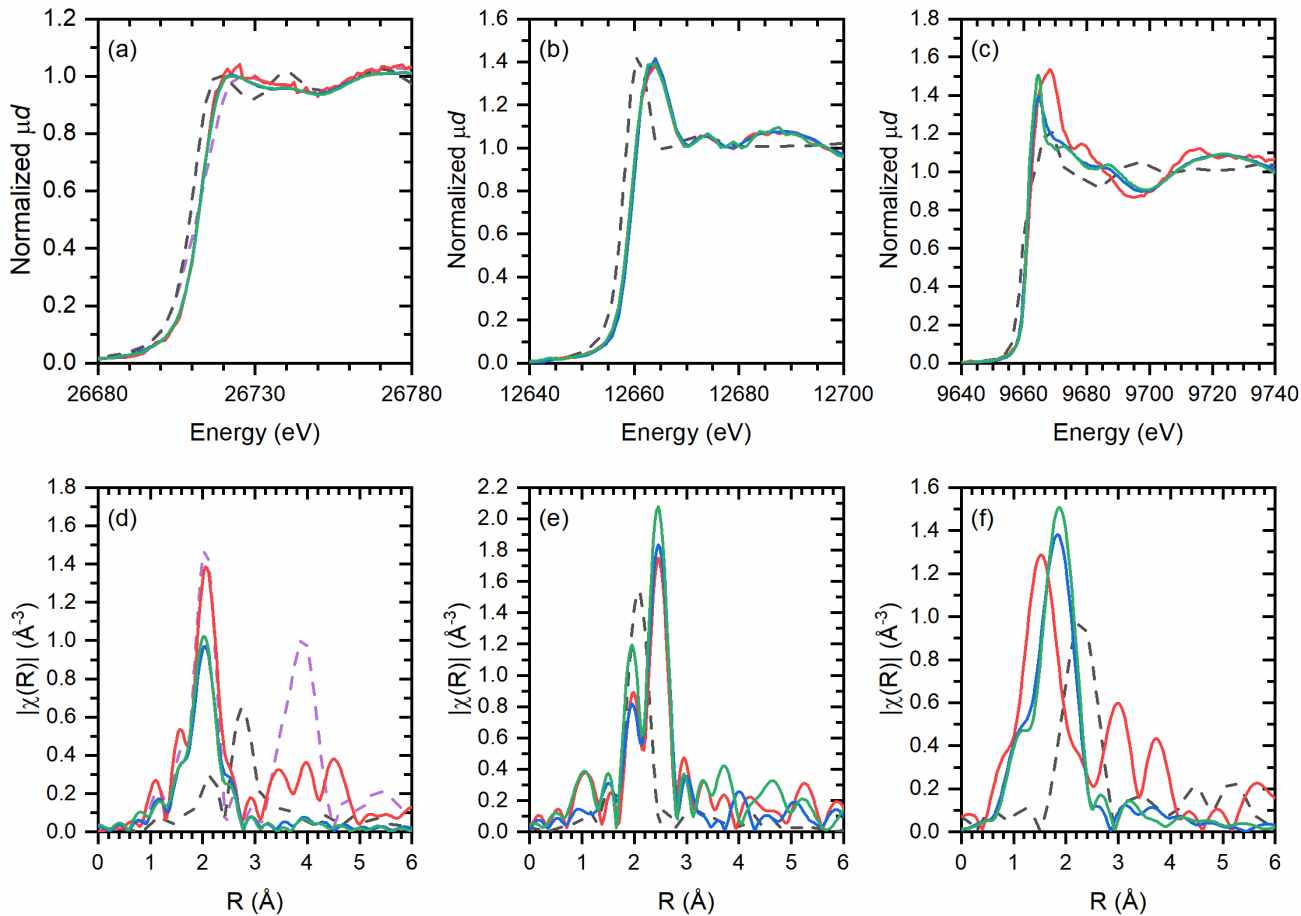

**Figure 3**. XANES spectra of CdSe/CdZnS NPLs with 3.5 (red lines), 4.5 (blue lines) and 5.5 ML (green lines) at Cd *K*-edge (a), Se *K*-edge (b) and Zn *K*-edge (c) and respective FT of EXAFS (d, e and f).

**Table 3**. Parameters extracted from the standard 1$^{st}$ and 2$^{nd}$ shell Fourier analyses of Cd *K*-edge.

| Name | Bonding | R (Å) | R err. (Å) | N | N err. | $\sigma^2$ | $\sigma^2$ err. |
|---|---|---|---|---|---|---|---|
| 3.5 ML CdSe/Cd$_x$Zn$_{1-x}$S | Cd-S | 2.3035 | 0.0991 | 1.663 | 0.810 | 0.0013 | 0.0053 |
| | Cd-Se | 2.5253 | 0.0517 | 4.108 | 4.681 | 0.0039 | 0.0056 |
| 4.5 ML CdSe/Cd$_x$Zn$_{1-x}$S | Cd-S | 2.4826 | 0.0468 | 2.367 | 0.867 | 0.0068 | 0.0087 |
| | Cd-Se | 2.6133 | 0.0550 | 0.720 | 1.106 | 0.0014 | 0.0056 |
| 5.5 ML CdSe/Cd$_x$Zn$_{1-x}$S | Cd-S | 2.4819 | 0.0473 | 2.502 | 0.977 | 0.0065 | 0.0103 |
| | Cd-Se | 2.5972 | 0.0660 | 0.580 | 1.237 | 0.0005 | 0.0070 |

The Se *K*-edge XANES spectra (Fig. 5B) show the shift towards higher energy with respect to the Se$^0$. Such a blue shift indicates higher oxidation state of the Se that can be described by a presence of Se-Cd bonding. This is also supported by FT of Se *K*-edge EXAFS (Fig. S10) that highlight a higher intensity peak at ~2.45 Å (phase uncorrected) and a smaller secondary peak at ~1.9 Å (phase uncorrected) associated with the Se–Cd bond. The 1$^{st}$ shell EXAFS analysis revealed interatomic distance 2.60-2.61 Å for all samples that is close to reported value in literature (2.63 Å).

**Table 4.** Parameters extracted from the standard 1st and 2nd shell FT analyses of Se *K*-edge.

| Name | Bonding | R (Å) | R err. (Å) | N | N err. | $\sigma^2$ | $\sigma^2$ err. |
|---|---|---|---|---|---|---|---|
| 3.5 ML CdSe/Cd$_x$Zn$_{1-x}$S | Se-Cd | 2.6093 | 0.0056 | 2.826 | 0.319 | 0.0037 | 0.0008 |
| 4.5 ML CdSe/Cd$_x$Zn$_{1-x}$S | Se-Cd | 2.6052 | 0.0051 | 3.296 | 0.313 | 0.0045 | 0.0007 |
| 5.5 ML CdSe/Cd$_x$Zn$_{1-x}$S | Se-Cd | 2.5968 | 0.0092 | 2.860 | 0.589 | 0.0026 | 0.0013 |

**Table 5.** Parameters extracted from the standard 1$^{st}$ shell Fourier analysis of Zn K-edge.

| Name | Bonding | R (Å) | R err. (Å) | N | N err. | $\sigma^2$ | $\sigma^2$ err. |
|---|---|---|---|---|---|---|---|
| 3.5 ML CdSe/Cd$_x$Zn$_{1-x}$S | Zn-O | 1.9914 | 0.0385 | 3.718 | 1.441 | 0.0038 | 0.0064 |
| 4.5 ML CdSe/Cd$_x$Zn$_{1-x}$S | Zn-S | 2.3394 | 0.0162 | 4.797 | 0.704 | 0.0107 | 0.0025 |
| 5.5 ML CdSe/Cd$_x$Zn$_{1-x}$S | Zn-S | 2.3573 | 0.0108 | 4.379 | 0.458 | 0.0081 | 0.0016 |

### 2.2.3. CdSe/CdZnS NPLs aging

The 4.5 ML CdSe/Cd$_x$Zn$_{1-x}$S NPLs are characterized by the highest value of QY (88-90%) making them the most prominent candidate for commercialization. We observed a perfect colloidal stability and unaltered PL properties for this structure. However, such structures must remain stable over time. Alteration of the NPLs pristine state during aging can be due to the ligand desorption, oxidation and aggregation/precipitation changes in the internal structure of the particles [44].

Thus, the 4.5 ML CdSe/Cd$_x$Zn$_{1-x}$S NPLs were measured after 2.5 and 6 months of storage inside a premise at ambient conditions with the scope of assessing the eventual diffusion of atoms through interfaces (the core-shell one and the external surface). From XAS measurements of Cd, Zn and Se *K*-edges we observed noticeable changes only for Cd local environment. The results for the Cd–S fit are presented in Figure 4 and in Table 6. The results for Zn (the Zn–S bonds) and Se are given in SI.

The decrease of coordination number (from 3.5 to 2.7) with aging might be followed by reorganization of the Cd centers. It is worth noting that Se preserves its state according to the XANES and EXAFS data. The following hypothesis might be drawn. Apparently, there are two populations of the Cd atoms: a smaller fraction is coordinated with Se that is stable over time while the majority is coordinated with S that is under external impact. It was found that, after 6 months of storage, the coordination number of Se around Cd atoms (3.5) slightly increases over that of Cd around Se atoms (2.7), evidencing that the NPLs may became Se-rich at the surface. A decrement in the Debye-Waller factor ($\sigma^2$) value (Table 6) can result from the weaker stress in the structure.

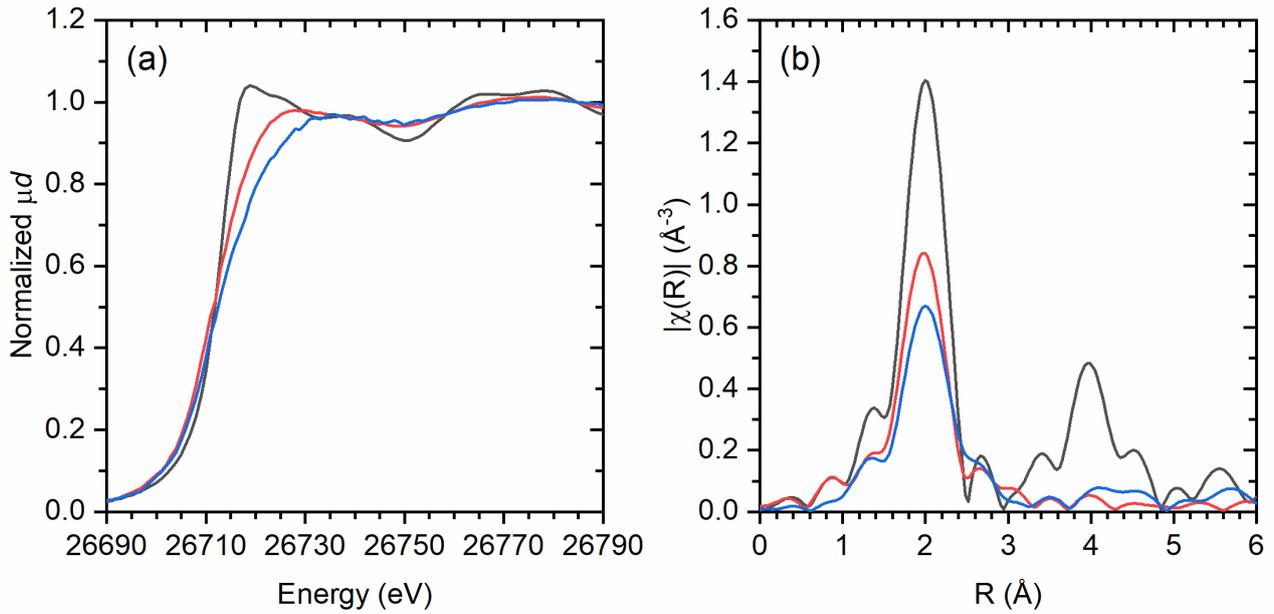

**Figure 4.** Normalized Cd $K$-edge XANES spectra (a) and FT of EXAFS (b) of the NPLs samples (4.5 ML CdSe/Cd$_x$Zn$_{1-x}$S 2.5 months (red line), 4.5 ML CdSe/Cd$_x$Zn$_{1-x}$S 6 months (blue line) and CdS (grey line).

**Table 6.** Fitting parameters derived from the EXAFS analysis.

| Sample | R (Å) | N | $\sigma^2$ (Å$^2$) |
|---|---|---|---|
| CdS | 2.524 ± 0.004 | 4 | 0.0033 ± 0.0006 |
| 4.5 ML CdSe/Cd$_x$Zn$_{1-x}$S 2.5 months | 2.491 ± 0.007 | 3.5 ± 0.3 | 0.0012 ± 0.0009 |
| 4.5 ML CdSe/Cd$_x$Zn$_{1-x}$S 6 months | 2.494 ± 0.019 | 2.7 ± 0.7 | 0.0006 ± 0.0026 |

We can assume that CdSe/Cd$_x$Zn$_{1-x}$S might be characterized as a stable structure, since after 1.5 years after the preparation changes were observed only for the surface Cd bonds. The spectral profiles revealing the coordination pattern of Cd in 4.5 ML CdSe/Cd$_x$Zn$_{1-x}$S NPLs appear very similar to those of a commercial CdSe/ZnS sample ("EviDots", Evident Technologies) (Fig. S12, Table S3). There is no difference observed in the Zn $K$-edge, which indicates that all changes in structure occur inside the particle: its core and the area near the interface, where the number of Cd-S decreased in favor of the Cd-Se bonding.

### 2.2.4. CdSe/CdS Core-Crown and CdSe/CdS/Cd$_x$Zn$_{1-x}$S NPLs

The CdSe crystal has a significant lattice parameter mismatch (12%) with the ZnS [45], which may cause stress at the interface when a core-shell structure is formed. On the other side the lattice mismatch between CdSe core with CdS crown is only about 3.9 %. This value is small enough for the epitaxial growth of CdS shell to proceed without alloying. The difference in bandgap values of these two cadmium chalcogenides allow to reach high QY of PL emission and ensures stability of the core [46]. Formation of the gradient in the form of interfacial alloy reduces the strain and the number of interfacial defects. It is well known that CdS shell has a finite confinement of the electron, and there is nearly no conduction band offset between CdSe and CdS at room temperature. The latter causes the delocalization of electron and its captivity by surface defects. XANES spectra of 3.5, 4.5 and 5.5 ML CdSe/CdS core-crown NPLs at the Cd K-edge (a) and the Se K-edge are shown in Figure 5a, b. One may observe high coordination of Cd for 3.5 ML CdSe/CdS NPLs (Table 7). The $\sigma^2$ value is also noticeably higher than that for 4.5 ML and 5.5 ML CdSe/CdS core-crown NPLs.

The ratio between the number of the Cd-S/Cd-Se bonds is 1.068 for 3.5 ML CdSe/CdS, 2.34 for 4.5 ML and 2.36 for 5.5 ML CdSe/CdS core-crown NPLs, respectively. It might correspond to the case when a fraction of the CdSe core surface remains unpassivated. That has been already observed for CdSe/CdS core-crown NPLs [47]. The FT Cd *K*-edge EXAFS (Figure 5d) contains a peak at ~2.1 Å for both the 4.5 ML and 5.5 ML CdSe/CdS core-crown NPLs, which corresponds to the dominant Cd−S coordination, whereas the peak at 2.3 Å for the 3.5 ML CdSe/CdS confirmed formation of the Cd-Se bond. There are no changes observed for Se *K*-edge (Table S4). FT of EXAFS data (Figure 5c) exhibit the Cd-Se bonds at 2.45 Å and the Se-Se bonds at 2.00 Å.

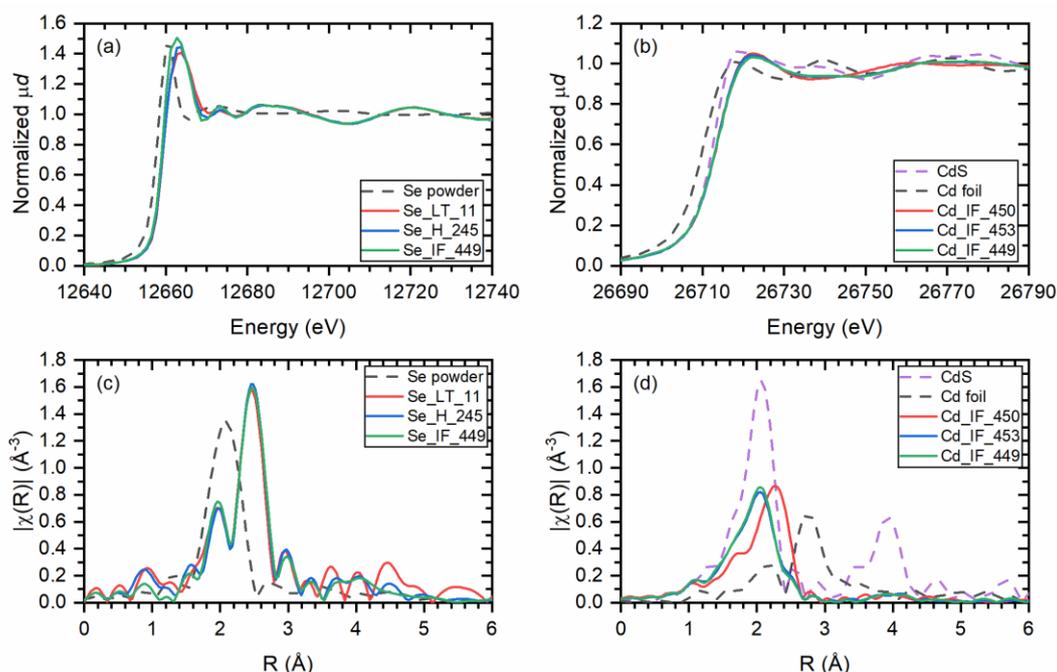

**Figure 5.** XANES spectra of 3.5, 4.5 and 5.5 ML CdSe/CdS core-crown NPLs at Cd K-edge (a) and Se K-edge (b) and the respective FT EXAFS (c and d).

**Table 7.** Parameters extracted from the standard 1st and 2nd shell FT analyses of Cd K-edge.

| Name | Bonding | R (Å) | R err. (Å) | N | N err. | $\sigma^2$ | $\sigma^2$ err. |
|---|---|---|---|---|---|---|---|
| 3.5 ML CdSe/CdS | Cd-S | 2.4283 | 0.0299 | 3.076 | 0.846 | 0.0247 | 0.0057 |
| | Cd-Se | 2.6172 | 0.0042 | 2.880 | 0.218 | 0.0061 | 0.0004 |
| 4.5 ML CdSe/CdS | Cd-S | 2.4620 | 0.0245 | 2.956 | 0.613 | 0.0124 | 0.0051 |
| | Cd-Se | 2.5968 | 0.0153 | 1.258 | 0.583 | 0.0046 | 0.0018 |
| 5.5 ML CdSe/CdS | Cd-S | 2.4586 | 0.0282 | 2.992 | 0.595 | 0.0125 | 0.0047 |
| | Cd-Se | 2.5905 | 0.0141 | 1.268 | 0.530 | 0.0044 | 0.0016 |

While the QY of the best CdSe/Cd$_x$Zn$_{1-x}$S NPLs solutions could reach 88-90%, the QY values of CdSe/CdS is significantly lower. For example, Norris group has successfully prepared CdSe/CdS NPLs by the hot-injection shell (HIS) approach, but a QY of 60was obtained as the highest possible value [8]. Both the QY and stability can be improved by forming the ZnS shell, however, the problem of low QY persists [48].

We prepared n ML (n = 3.5, 4.5, 5.5) CdSe/CdS/Cd$_x$Zn$_{1-x}$S NPLs, and found out that QY of all CdSe/CdS/Cd$_x$Zn$_{1-x}$S structures was dramatically lower those of CdSe/Cd$_x$Zn$_{1-x}$S. The QY values were 30% and 24.5% for 5.5 ML CdSe/Cd$_x$Zn$_{1-x}$S and 5.5 ML CdSe/CdS/Cd$_x$Zn$_{1-x}$S, respectively. They were 89% and 55.8% for 4.5 ML CdSe/Cd$_x$Zn$_{1-x}$S and 4.5 ML CdSe/CdS/Cd$_x$Zn$_{1-x}$S, accordingly. Finally, they dropped to just 3% and 1.9% for 3.5 ML CdSe/Cd$_x$Zn$_{1-x}$S and 3.5 ML CdSe/CdS/Cd$_x$Zn$_{1-x}$S, respectively.

Figure 6 shows XANES spectra of Cd, Se and Zn *K*-edges and FT profiles for the 3.5, 4.5 and 5.5 ML CdSe/CdS/Cd$_x$Zn$_{1-x}$S NPLs. In the case of 3.5 ML CdSe/CdS/Cd$_x$Zn$_{1-x}$S NPLs the coordination number of Cd-Se was equal to 4.7 that is close to the bulk CdSe. The ratio of the number of the Cd-Se/Cd-S bonds is 1.58, which suggests beneficial formation of CdSe. A nearly epitaxial growth of the shell with minimal mixing and gradient formation can be expected in this case.

Formation of the Cd$_x$Zn$_{1-x}$S shell over the CdSe/CdS core-crown leads to increasing of interatomic distances for Cd-S bonds in the 3.5 ML CdSe/CdS/Cd$_x$Zn$_{1-x}$S NPLs in comparison with CdSe/Cd$_x$Zn$_{1-x}$S (from 2.3035 Å to 2.3898 Å), whereas for the 4.5 and 5.5 ML CdSe/CdS/Cd$_x$Zn$_{1-x}$S NPLs interatomic distances are similar. The Cd-Se bond distance is also increased after the shell formation over CdSe/CdS core-crown for the 3.5 and 4.5 ML CdSe/CdS/Cd$_x$Zn$_{1-x}$S NPLs, whereas for the 5.5 ML CdSe/CdS/Cd$_x$Zn$_{1-x}$S some compression has been observed (due to shortening of the bond length from 2.5972 to 2.5803 Å).

The $\sigma^2$ value for the 4.5 and 5.5 ML CdSe/CdS/Cd$_x$Zn$_{1-x}$S NPLs is lower than that for CdSe/Cd$_x$Zn$_{1-x}$S. This is probably due to a larger degree of structural disorder and a stronger stress at the interface in CdSe/Cd$_x$Zn$_{1-x}$S. Comparison of the CdSe/CdS core-crown and CdSe/CdS/Cd$_x$Zn$_{1-x}$S NPLs show a significant reduction of the Cd-S bonds after the formation of a shell (from 2.4283 Å to 2.3898 Å) in the 3.5 ML system and increase in the length for 4.5 ML (from 2.4620 Å to 2.4909Å) and 5.5 ML (from 2.4586 Å to 2.4773Å). The $\sigma^2$ value lowered for the 3.5 ML CdSe/CdS/Cd$_x$Zn$_{1-x}$S NPLs that also could be indicative of an incomplete coverage of 3.5 ML CdSe with CdS crown in CdSe/CdS NPLs, and formation of the intermediate CdS shell before the Cd$_x$Zn$_{1-x}$S shell growth.

Coordination number of Se-Cd bond slightly increased with the increase of number of layers (Fig. S13, Table S4), while Debye-Waller factor keeps constant at the same time the reduction of the Se-Cd bond of 4.5 ML CdSe core with respect to the 4.5 ML CdSe/CdS/Cd$_x$Zn$_{1-x}$S NPLs was observed. The effect of the intermediate CdS layer on the structure of ZnS shell (Zn *K*-edge, Table 9) was clearly observed in the case of 3.5 ML-based systems. The Zn-O bonds we detected in the 3 ML CdSe/Cd$_x$Zn$_{1-x}$S, but after intermediate formation of the CdS layer just the Zn-S bonds were found with increasing coordination of Zn (from 3.718 to 5.368). The Zn-S bond length increases in the row: 3.5 ML < 5.5 ML < 4.5 ML CdSe/CdS/Cd$_x$Zn$_{1-x}$S. Instead, the $\sigma^2$ value decreases and it reached minimum for the 4.5 ML CdSe/CdS/Cd$_x$Zn$_{1-x}$S NPLs at Zn coordination of 4 (which is close to that in bulk ZnS). It could be concluded that for the 4.5 ML CdSe/CdS formation of Cd$_x$Zn$_{1-x}$S shell leads to ordering of the system resulting in higher QY in comparison to both the 3.5 and 5.5 ML CdSe/CdS/Cd$_x$Zn$_{1-x}$S. However, further investigation of the core-crown-shell structures with a thicker CdS shell is needed.

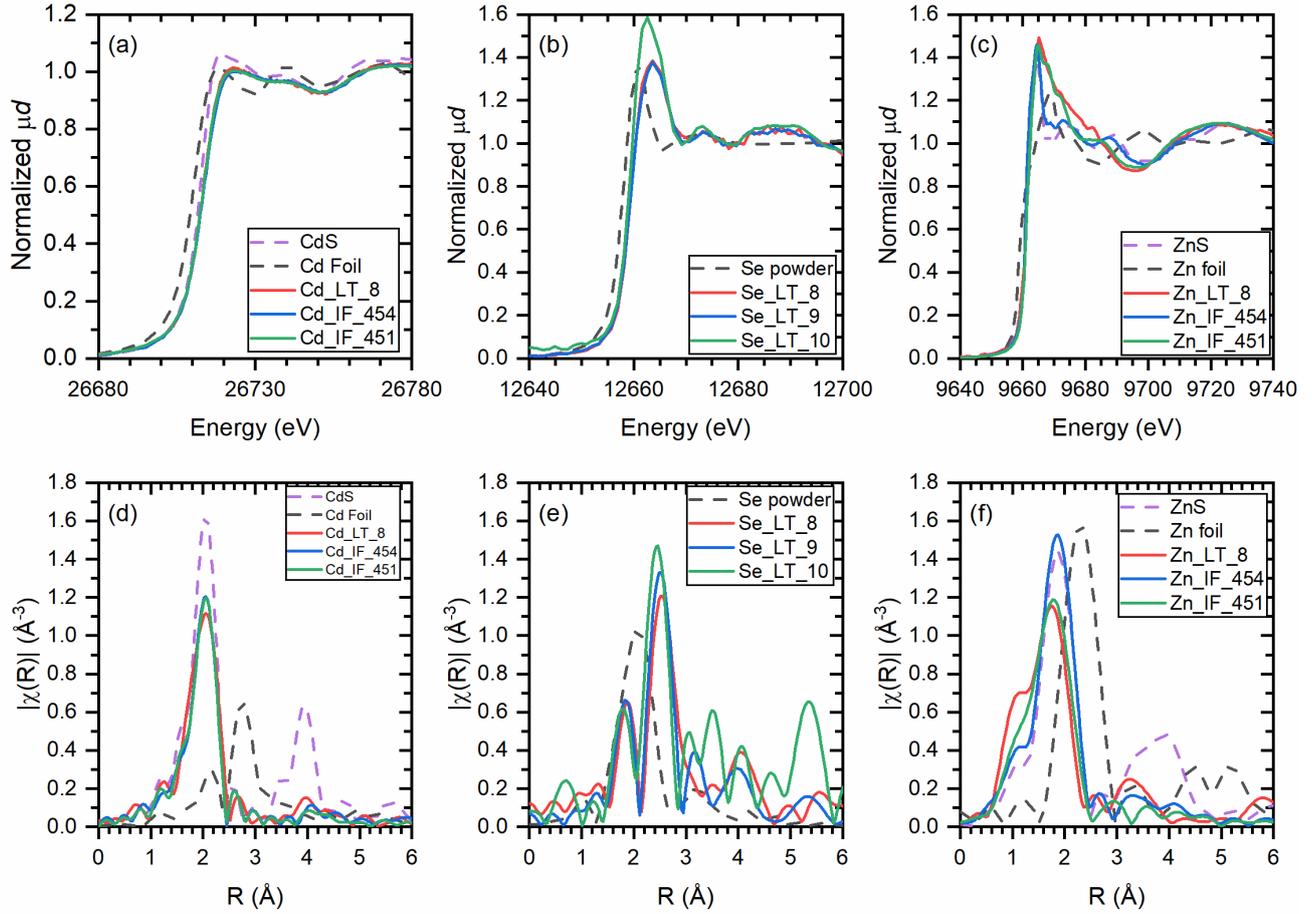

**Figure 6**. XANES spectra of 3.5, 4.5 and 5.5 ML CdSe/CdS/Cd$_x$Zn$_{1-x}$S NPLs at Cd *K*-edge (a), Se *K*-edge (b) and Zn *K*-edge (c) and respective Fourier transformed of EXAFS (d, e and f).

**Table 8**. Parameters extracted from the standard 1$^{st}$ and 2$^{nd}$ shell FT analyses of Cd *K*-edge data for 3, 4 and 5 ML CdSe/CdS/Cd$_x$Zn$_{1-x}$S NPLs

| Name | Bonding | R (Å) | R err. (Å) | N | N err. | $\sigma^2$ | $\sigma^2$ err. |
|---|---|---|---|---|---|---|---|
| 3.5 ML | Cd-S | 2.3898 | 0.0836 | 2.988 | 3.382 | 0.0021 | 0.0069 |
| CdSe/CdS/Cd$_x$Zn$_{1-x}$S | Cd-Se | 2.5661 | 0.0785 | 4.729 | 6.963 | 0.0060 | 0.0093 |
| 4.5 ML | Cd-S | 2.4909 | 0.0247 | 2.911 | 0.432 | 0.0037 | 0.0028 |
| CdSe/CdS/Cd$_x$Zn$_{1-x}$S | Cd-Se | 2.6544 | 0.2929 | 0.141 | 0.842 | 0.0028 | 0.0330 |
| 5.5 ML | Cd-S | 2.4773 | 0.0507 | 2.896 | 0.450 | 0.0041 | 0.0095 |
| CdSe/CdS/Cd$_x$Zn$_{1-x}$S | Cd-Se | 2.5803 | 0.0592 | 0.370 | 1.122 | 0.0003 | 0.0106 |

Table 9. Parameters extracted from the standard 1st shell FT analyses of Zn K-edge for 3.5, 4.5 and 5.5 ML CdSe/CdS/Cd$_x$Zn$_{1-x}$S NPLs

| Name | Bonding | R (Å) | R err. (Å) | N | N err. | $\sigma^2$ | $\sigma^2$ err. |
|---|---|---|---|---|---|---|---|
| 3.5 ML CdSe/CdS/Cd$_x$Zn$_{1-x}$S | Zn-S | 2.2706 | 0.0292 | 5.368 | 1.408 | 0.0160 | 0.0047 |
| 4.5 ML CdSe/CdS/Cd$_x$Zn$_{1-x}$S | Zn-S | 2.3516 | 0.0114 | 4.039 | 0.500 | 0.0074 | 0.0018 |
| 5.5 ML CdSe/CdS/Cd$_x$Zn$_{1-x}$S | Zn-S | 2.3121 | 0.0223 | 5.159 | 1.117 | 0.0144 | 0.0036 |

## CONCLUSIONS

This work is the first XAS systematic study of the core-shell interface in CdSe/Cd$_{1-x}$Zn$_x$S core-shell NPLs. It provides important insights into preferential bonding in that internal region from different views as the CdSe core size varies from 3.5 to 5.5 ML and the Cd, Zn, and Se were probed at K edge. This allows to build a comprehensive picture of element arrangement in the gradient shells. This work also provides critical insights into time stability of the core-shell system, demonstrating that it may stay unaltered for many months at normal conditions. Of course, such delicate study has own limits, too. The picture becomes too complex for the analysis relying on XAS data only for the core-crown-shell samples, due to their structural complexity. The dominating bonds at the interfaces together with other structural information provide useful insights into structural origin of observed optical phenomena as the energy shift of PL emission when tuning the thickness of core and adding a shell.

## ACKNOWLEDGMENT

The authors gratefully acknowledge the support of XAFS/XRF Beamline of Synchrotron-light for Experimental Science and Applications in the Middle East (SESAME), Jordan, TENMAK for the financial support. The authors also acknowledge the financial support in part from TUBITAK 119N343, 20AG001, 121N395, and 121C266. H.V.D. acknowledges the support from TUBA and TUBITAK 2247-A National Leader Researchers Program (121C266). T.L. acknowledges the financial support from TUBITAK 2221 Scientist support program with guest or academic permission.

## ASSOCIATED CONTENT

Supporting Information. The following files are available free of charge.

**Author Contributions**

The manuscript was written through the contributions of all authors. All authors have approved the final version of the manuscript. LT: conceptualization, synthesis, synchrotron measurements, lab-scale measurements, methodology, data analysis, data curation, validation, visualization, writing original draft, review, and editing. OU: synchrotron measurements, data analysis, writing original draft, review, and editing, MH: synchrotron measurements, data analysis, FI, BC: synthesis/sample preparation, synchrotron measurements, and report review. AB: lab-scale measurements, data analysis, original draft, review, and editing. HVD: conceptualization, data analysis, writing original draft, review, and editing.

# Supporting Information

# Unveiling the local elemental arrangements across the interfaces inside CdSe/Cd$_{1-x}$Zn$_x$S core-shell and CdSe/CdS/Cd$_{1-x}$Zn$_x$S core-crown-shell quantum wells

*Tatiana Lastovina[1], Oleg Usoltsev[2], Furkan Işik[1], Messaoud Harfouche[3], Andriy Budnyk[1], Betül Canımkurbey[1], and Hilmi Volkan Demir[1, 4, *]*

[1] Department of Electrical and Electronics Engineering, Department of Physics, UNAM − Institute of Materials Science and Nanotechnology and The National Nanotechnology Research Center, Bilkent University, Ankara 06800, Turkey.

[2] ALBA Synchrotron, Carrer de la Llum, 2, 26, 08290 Cerdanyola del Vallès, Barcelona, Spain.

[3] Synchrotron-light for Experimental Science and Applications in the Middle East (SESAME), P.O. Box 7, Allan 19252, Jordan.

[4] Luminous! Center of Excellence for Semiconductor Lighting and Displays, School of Electrical and Electronic Engineering, Division of Physics and Applied Physics, School of Physical and Mathematical Sciences, School of Materials Science and Engineering, Nanyang Technological University, Singapore 639798, Singapore.

# 1. Synthetic procedures

## 1.1. Chemicals

The reagents, 1-octadecene ($C_{18}H_{36}$, 1-ODE, 90%, Aldrich), oleic acid (OA, 90%, Aldrich), oleylamine (OAm, technical grade, 70%, Aldrich), selenium (100 mesh powder, 99.99%, Aldrich), sulfur flakes (>99.99%, Aldrich), cadmium acetate dihydrate ($Cd(OAc)_2 \cdot 2H_2O$, Aldrich), cadmium acetate anhydrous ($Cd(OAc)_2$, Aldrich), zinc acetate ($Zn(OAc)_2$, anhydrous, 99.98%, Alfa-Aesar), sodium myristate ($CH_3(CH_2)_{12}COONa$, ≥99%, Aldrich), cadmium nitrate tetrahydrate ($Cd(NO_3)_2 \cdot 4H_2O$, 99.997% trace metals basis, Aldrich), ethanol ($C_2H_5OH$, absolute, ≥99.8%, Aldrich), methanol ($CH_3OH$, absolute, acetone free, Aldrich), acetone ($CH_3COCH_3$, suitable for HPLC, ≥99.9%, Aldrich) were used as received. Octane-1-thiol ($CH_3(CH_2)_7SH$) and OAm were dried before use. High purity (99.999%) Ar was used for setting an inert atmosphere.

## 1.2. Synthesis of 3.5 ML CdSe

*Preparation of the Se-stock solution.* 300 mg of Se was added to 25 mL of 1-ODE, and a three-neck flask with a condenser and a magnetic stirring was assembled. Suspension was hold under vacuum at 80 °C, which it was degassed with Ar. The temperature was increased up to 240 °C until the color of solution was turned yellow.

*Preparation of the 3.5 ML CdSe colloidal solution.* 8 mL of as-prepared Se-stock solution was mixed with 40 mL of 1-ODE, 860 mg of $(CH_3COO)_2Cd \cdot 2H_2O$ and 1.4 mL of OA. Solution was degassed with Ar and heated up to 250 °C. After that the reaction was quenched using the water bath. After cooling the mixture to room temperature, CQWs were separated by centrifugation, washed, and re-dispersed in hexane.

## 1.3. Synthesis of 4.5 ML CdSe

*Synthesis of cadmium myristate.* Cadmium myristate was synthesized following a previously reported protocol with slight modifications [R[1]]. 6.26 g of sodium myristate and 2.46 g of cadmium nitrate tetrahydrate were separately dissolved in 500 and 80 mL of methanol, respectively. Then, the solutions were mixed and stirred vigorously for 5 h with a magnetic stirrer at room temperature. After completion of the reaction, the white precipitate (cadmium myristate) was filtered out using a Büchner funnel. The product was washed several times with methanol and vacuum-dried overnight.

*Preparation of the 4.5 ML CdSe colloidal solution.* 341 mg of pre-prepared cadmium myristate and selenium powder were mixed with 30 mL 1-ODE in a three-neck flask. The mixture under vigorous stirring was degassed at 90 °C under the vacuum. An inert gas was submitted into the flask, and the temperature was set to 240 °C. When the color of the solution turned deep orange (at ~190-200 °C), 120 mg of cadmium acetate dihydrate was added. The mixture's temperature was kept at 240 °C for 8 min favoring the lateral growth of CQWs. After quenching reaction using water bath, at 200 °C, 1 mL of OA was added. After cooling the mixture to room temperature, CQWs were separated by centrifugation, washed, and re-suspended in hexane.

### 1.4. Synthesis of 5.5 ML CdSe

650 mg of cadmium myristate was mixed with 56 mL of 1-ODE in a three-neck flask and held under the vacuum for 10 min. Subsequently, the temperature was increased up to 90 °C with degas, and then up 250 °C. When the temperature reached 250 °C, solution containing 49 mg of Se in 4 mL of 1-ODE was rapidly injected, and 258 mg of $(CH_3COO)_2Cd \cdot 2H_2O$ 60 s later. The mixture was hold at 250 °C 1 h, cooled with water bath. At 200 °C 2 mL of OA was added. After cooling the mixture to room temperature, CQWs were separated by centrifugation, washed, and re-dispersed in hexane.

### 1.5. Synthesis of nML CdSe/CdS (n=3.5, 4.5, 5.5) core-crown NPLs

*Cd precursor synthesis.* 4 mL of 1-ODE, 960 mg of $CH_3COO)_2Cd \cdot 2H_2O$ and 690 mg of OA were mixed in a flask and sonicated for 1-2 h. Then, the mixture was stirred at 200 °C. Water in ultrasonic bath was changed to boiling water. The mixture was placed on a hot plate for 7 min and stirrer for 3 min. The procedure was repeated by decreasing the temperature gradually 10 °C for each 10 min till 150 °C, while it was on the hot plate loosen the cap of the flask. When 140 °C was reached the solution was transferred to a clean flask and the remaining jelly precipitate on the bottom of the flask was discarded. After the synthesis the suspension was kept under continuous stirring.

*Synthesis of nML CdSe/CdS core-crown NPLs.* As-prepared nML (x = 3.5, 4.5, 5.5) core CdSe CQWs were consequently washed in ethanol and hexane and collected by centrifugation. Finally, they were dispersed in 15 mL of ODE and transferred to a three-necked glass cell. 300 µL of oleic acid was added. After stirring under a vacuum until bubbling, the temperature has risen to 90-95 °C. At this point, a certain amount of as-prepared Cd precursor and S-ODE solutions were added.

### 1.6. Synthesis of nML CdSe/CdZnS (n = 3.5, 4.5, 5.5)

A certain amount of the nML (x = 3.5, 4.5, 5.5) CdSe CQWs was washed at least three times in ethanol and re-dispersed in 1 mL of hexane. Solutions containing CdSe CQWs, 22 mg of $Cd(CH_3COO)_2$, 55 mg of $Zn(CH_3COO)_2$, 750 µL of OA, and 7.5 mL of ODE were prepared and vacuumed at room temperate for 60 min. Then the temperature was increased to 80 °C, and the mixture was left stirring for 45 min.

The solutions of 750 µL of OAm containing 8 mL ODE and 140 µL of octane-1-thiol were prepared in a glove box. The main mixture was bubbled with Ar, OAm solution was added, and the temperature was set to 300 °C. Injection of octane-1-thiol solution started at ~161 °C with the rate of 10 mL/h until 240 °C was reached. It was continued at 4 mL/h afterwards. After the injection, the mixture was slightly cooled and the reaction was quenched by immersing the vessel in the water bath.

### 1.7. Synthesis of nML CdSe/CdS/CdZnS (n = 3.5, 4.5, 5.5)

A certain amount of the nML (n = 3.5, 4.5, 5.5) CdSe/CdS core-crown CQWs was washed at least three times in ethanol and redispersed in 1 mL of hexane. Solutions containing CdSe/CdS CQWs, 22 mg of $Cd(CH_3COO)_2$, 55 mg of $Zn(CH_3COO)_2$, 750 µL of OA, and 7.5 mL of ODE were prepared and vacuumed at room temperate for 60 min. Then the temperature was increased to 80 °C, and the mixture was left stirring for 45 min.

The solutions of 750 µl of OAm containing 8 ml ODE and 140 µl of octane-1-thiol were prepared in a glove box. The main mixture was bubbled with Ar, OAm solution was added, and the temperature was set to 300 °C. Injection of octane-1-thiol solution started at ~161 °C with the rate of 10 ml/h until 240 °C was reached and continued at 4 ml/h afterward. After the injection, the mixture was cooled and the reaction was quenched by immersing the vessel in the water bath.

## 2. Characterizations techniques

### 2.1. X-Ray diffraction

X-ray diffraction (XRD) scans were recorded on a X'Pert Panalytical diffractometer operating at 45 kV and 40 mA (Cu $K_\alpha$ irradiation, $\lambda=1.542$ Å). Data collection covered a 2θ range from 10° to 90°, utilizing a step increment of 0.01° and a dwell time of 150 s per step. For the measurements samples were redispersed from chloroform solutions, dropped onto the surface of a glass slide and naturally dried.

### 2.2. Transmission electron microscopy

A TEM imaging, high-resolution TEM (HRTEM) imaging, Energy Dispersive X-ray Spectroscopy (EDS) and selected area electron diffraction (SAED) were performed on HITACHI HF5000 200 kV (S)TEM in n2STAR center of Koç University (Istanbul, Türkiye). Samples for the analysis were prepared by dropping a dilute hexane solution of NPLs onto the ultrathin carbon-coated copper grids. The excess amount was washed in acetone and sample was left drying naturally. TEM images were processed with Digimizer.

### 2.3. Fourier-transformed infrared spectroscopy

Infrared spectra were recorded at Tensor 37 Bruker FTIR spectrometer equipped with ATR (attenuated total reflectance) accessory. The sample was drop-casted on a diamond crystal and left accommodate before the measurement. Spectra were recorded in the 400-4000 $cm^{-1}$ interval with 64 acquisitions with 4 $cm^{-1}$ resolution.

### 2.4. Photoluminescence emission spectroscopy

Photoluminescence (PL) spectra were measured on an Agilent Cary Eclipse fluorescent spectrometer in solutions filled in a 10 mm standard quartz cuvette. The excitation and emission light intensity were adjusted to remain in scale by varying the width of slits.

## 3. XAFS measurements

X-ray absorption fine structure (XAFS) data processing is a critical aspect of extracting meaningful structural information from X-ray absorption spectroscopy experiments. It is a powerful technique used to investigate the local atomic environment of materials and extract structural parameters includes interatomic distances (R), coordination number (N), Debye-Waller parameter ($\sigma 2$) and oxidation state.

XAFS data was collected at the Synchrotron-Light for Experimental Science and Applications in the Middle East (SESAME) and analyzed using Demeter software package. The collected spectra were calibrated using reference spectra collected simultaneously with the the first sample using the second and third ionization chambers. The chosen references were Zn foil, Se and CdO. Each spectrum was normalized by subtracting the pre-edge function and divided on the post-edge function. The experiment involved recording the X-ray absorption spectra as a function of the incident X-ray energy.

Extended X-ray absorption fine structure (EXAFS) spectrum was converted into k-space, where k represents the wave vector of the photoelectron. This transformation was achieved by subtracting the energy at each data point from the absorption edge energy and converting the result into units of Å$^{-1}$. The k-space data was used for further analysis. The heart of EXAFS data analysis is the Fourier transform (FT). The FT transforms the oscillatory EXAFS signal in k-space into real space, generating an $\chi(R)$ function that is similar to the radial distribution function. The FT reveals information about the distances R, coordination numbers, and thermal vibrations of neighboring atoms around the absorbing atom.

XAFS measurement was performed on the XAFS/XRF beamline of the SESAME operated at 2.5 GeV in the "decay" mode with a maximum electron current of 250 mA. XAFS spectra of the CdSe/Cd$_{1-x}$Zn$_x$S nanocrystals were acquired in the both transmission and fluorescent modes in the spectral range of Zn K-edge (9659 eV) and in fluorescence mode at Se K-edge (12658 eV) and Cd K-edge (26711 eV) at room temperature, using silicon drift detectors (SDD), KETEK GmbH. X-ray beam intensity before and after the sample was measured by the ionization chambers filled with an optimal mixture of noble gases at a total pressure of 1.0 bar and data of the samples and references were acquired from the signals measured at the ion chambers subsequently amplified by Stanford picoammeters and digitalized by a voltage to frequency converter, using a double-crystal Si (111) monochromator. The energy of the monochromator was calibrated with reference Zn, Se and Cd foils at their corresponding K-edges. The sample was prepared in small pellet form by the impregnation of the polyvinylpyrrolidone (PVP) powder with colloidal solutions of NPLs, drying and pressing a homogeneous mixture. A minimum of three scans were collected, averaged and normalized to improve the signal-to-noise ratio. To obtain quantitative structural information, the processed EXAFS spectrum was fitted using theoretical models of Artemis software. These models incorporate various structural parameters, including bond distances, coordination numbers, and Debye-Waller factors. The fitting process adjusts these parameters to achieve the best agreement between the experimental and model spectra, providing a precise characterization of the local atomic structure. XANES theoretical calculations were carried out using finite difference method available FDMNES software package. The calculation radius was selected to be 6.0 Å. The structures of Cd-Se from open crystallography database were used and modified to create surface structures with different number of monolayers.

## 4. Laboratory characterization results

### 4.1. X-ray diffraction

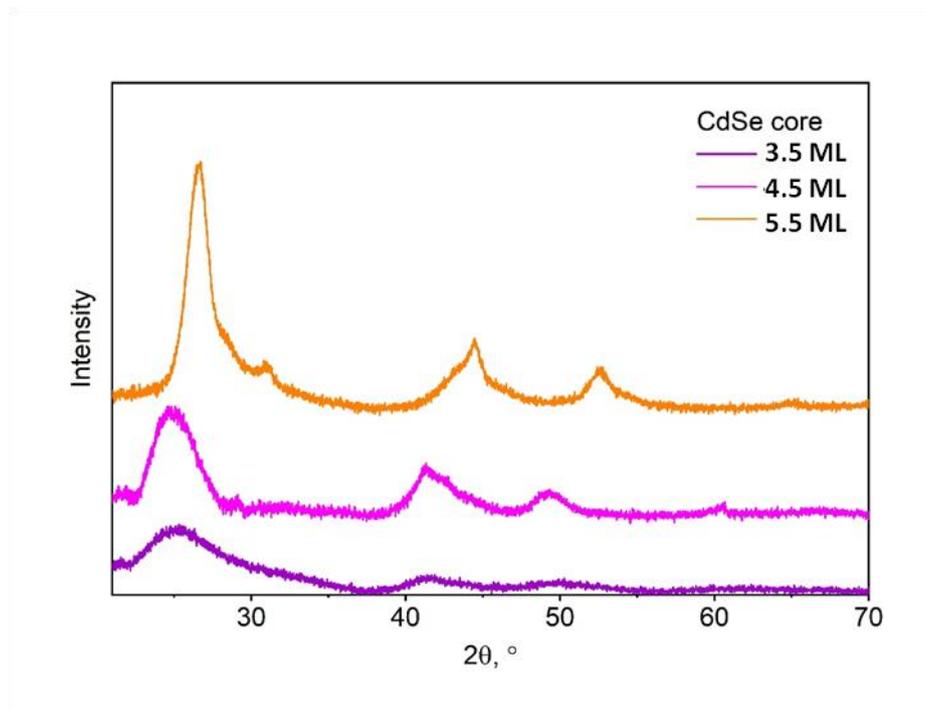

**Fig. S1.** XRD patterns of CdSe core samples of 3.5, 4.5 and 5.5 ML.

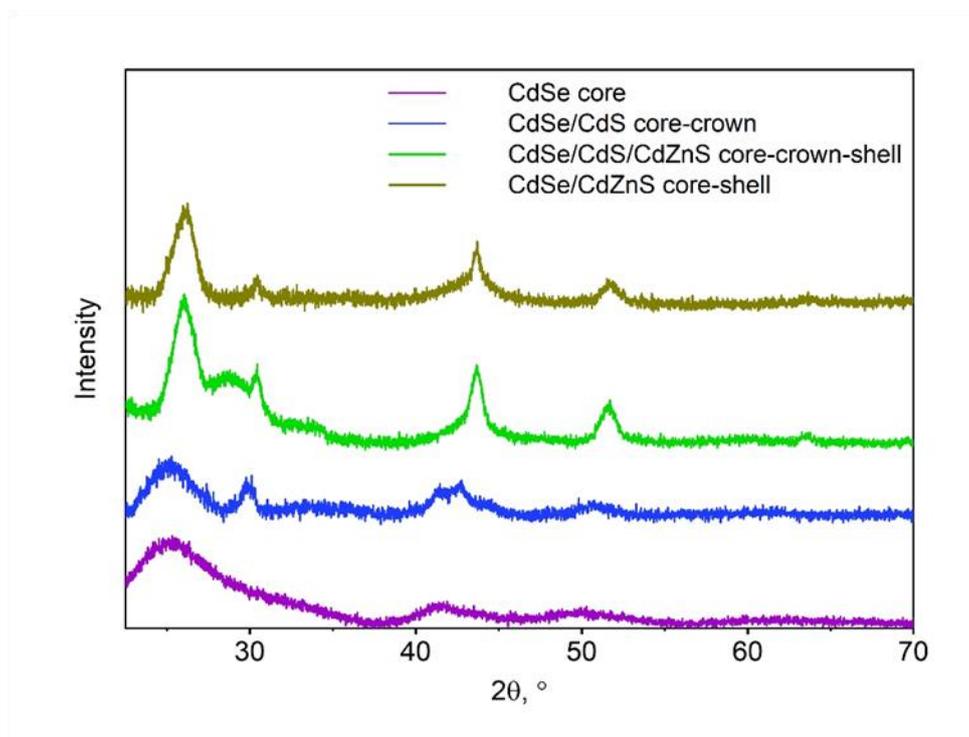

**Fig. S2.** XRD patterns of 3.5 ML samples with different heterostructure.

## 4.2. Transmission electron microscopy (TEM)

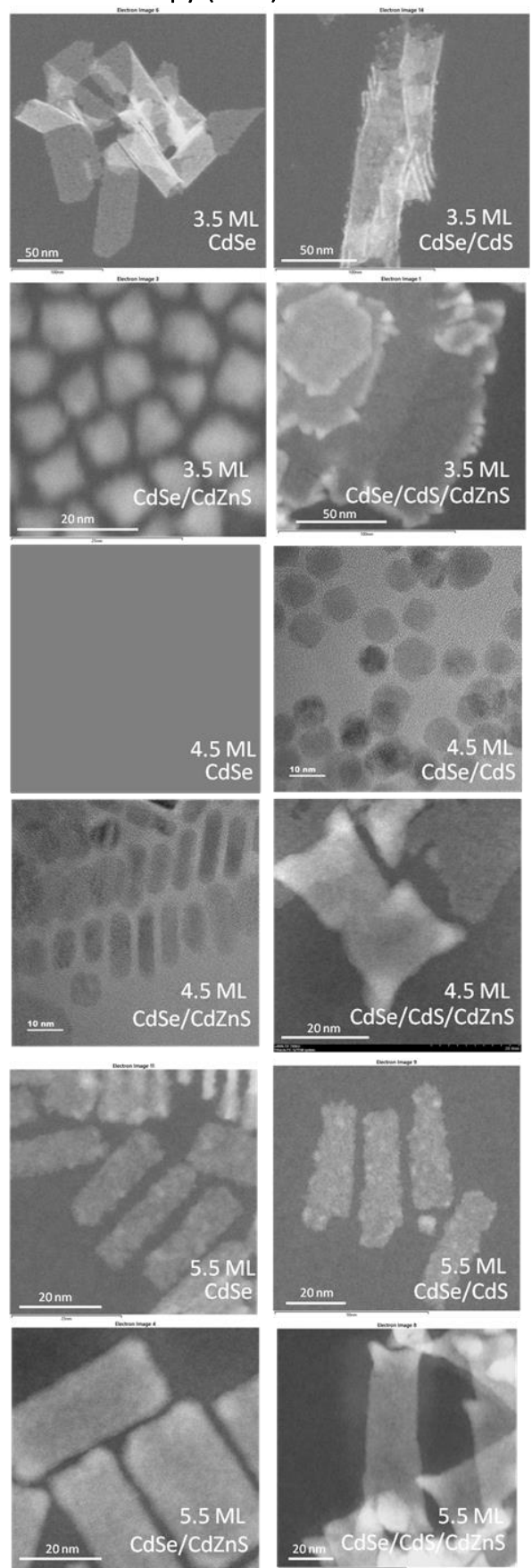

**Fig. S3.** TEM images of 3.5 ML (above), 4.5 ML (center) and 5.5 ML (below) NPLs.

**Table S1.** Mean size values of NPLs based on the graphical analysis of the TEM images.

| NPLs | Length, nm | SD$_L$, nm | Width, nm | SD$_W$, nm |
|---|---|---|---|---|
| **3.5 ML** | | | | |
| CdSe/CdS/CdZnS | 18.3 | 7.5 | 15.7 | 4.5 |
| CdSe/CdZnS | 7.7 | 0.9 | 5.6 | 0.9 |
| **4.5 ML** | | | | |
| CdSe/CdS | 10.3 | | 1.0 | |
| CdSe/CdS/CdZnS | 29.8 | 4.6 | 20.4 | 1.5 |
| CdSe/CdZnS | 16.1 | 2.9 | 5.2 | 0.9 |
| **5.5 ML** | | | | |
| CdSe | 28.6 | 3.5 | 7.6 | 1.2 |
| CdSe/CdS | 40.2 | 4.9 | 11.7 | 2.9 |
| CdSe/CdS/CdZnS | 60.3 | 7.8 | 22.8 | 2.2 |
| CdSe/CdZnS | 41.1 | 3.7 | 18.2 | 1.8 |

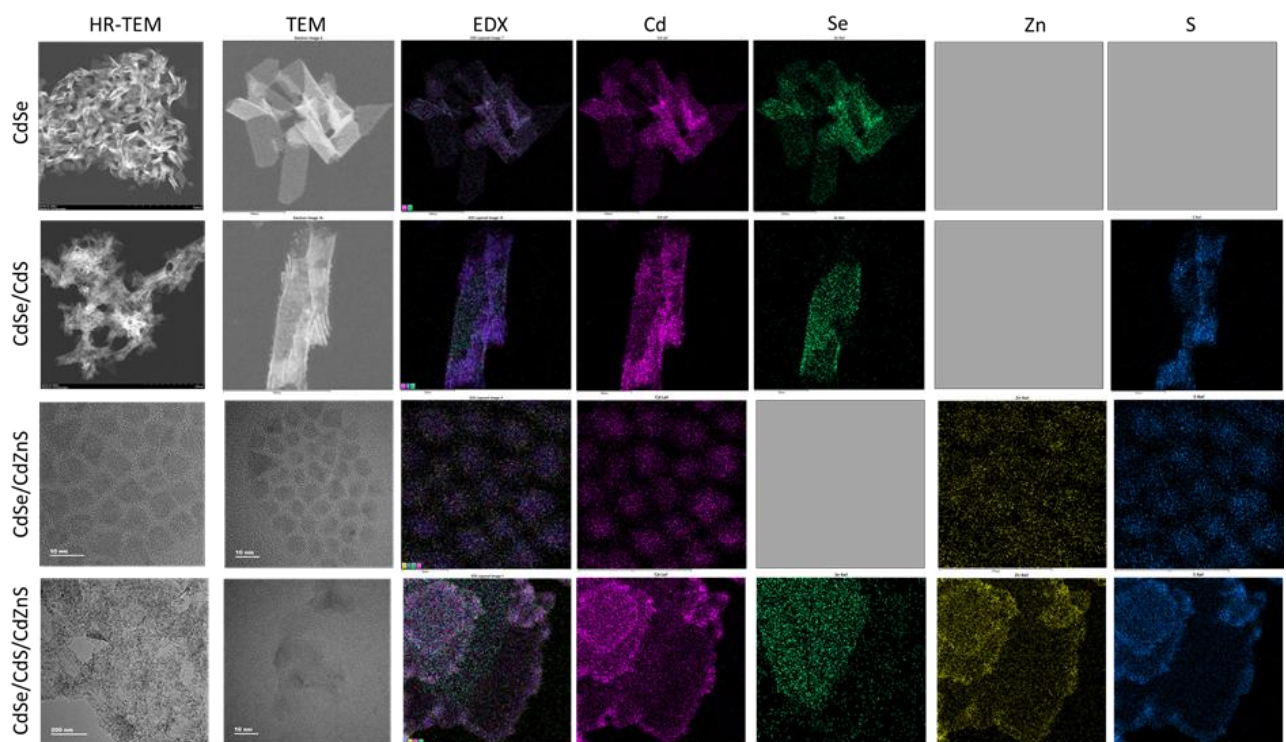

**Fig. S3-1.** STEM-EDX elemental analysis of 3.5 ML NPLs.

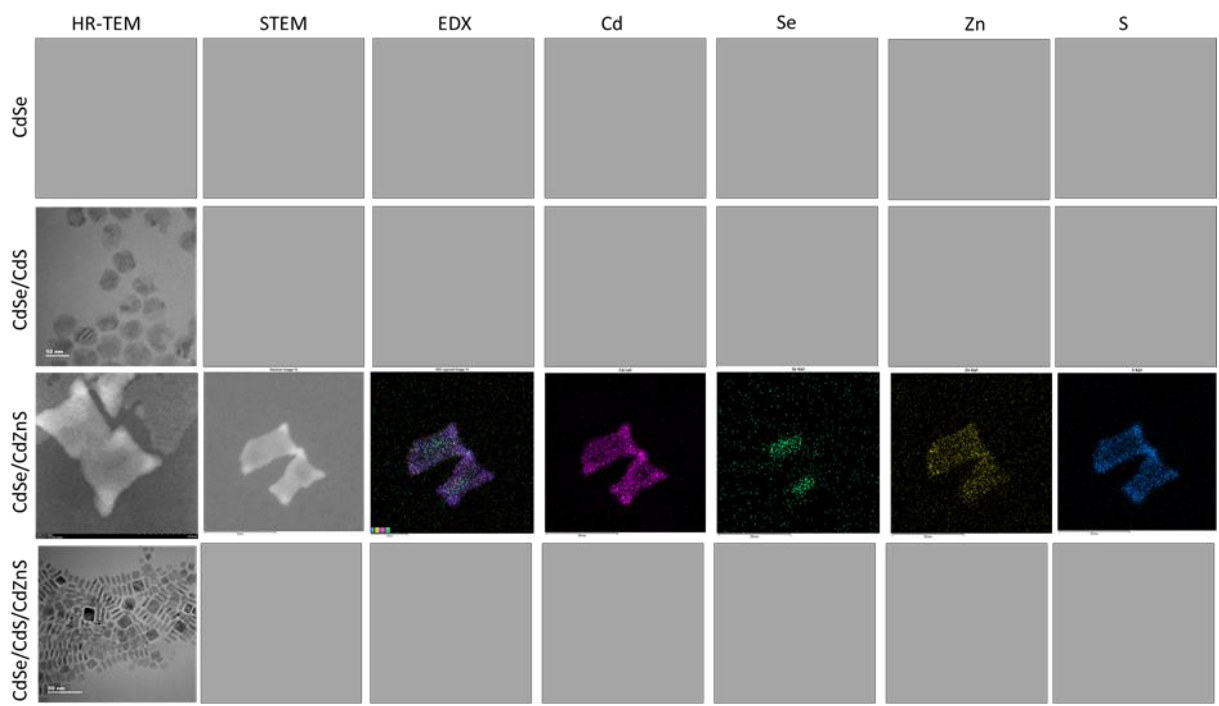

**Fig. S3-2.** STEM-EDX elemental analysis of 4.5 ML NPLs.

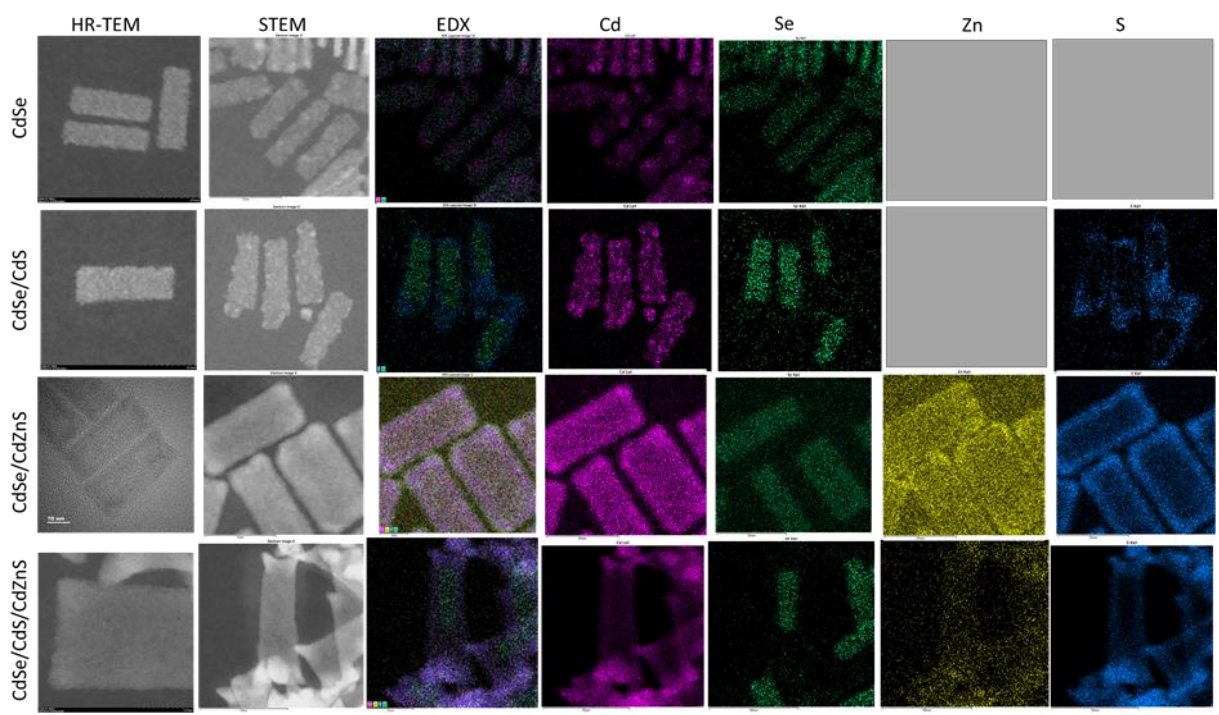

**Fig. S3-3.** STEM-EDX elemental analysis of 5.5 ML NPLs.

## 4.3. Photoluminescence spectroscopy

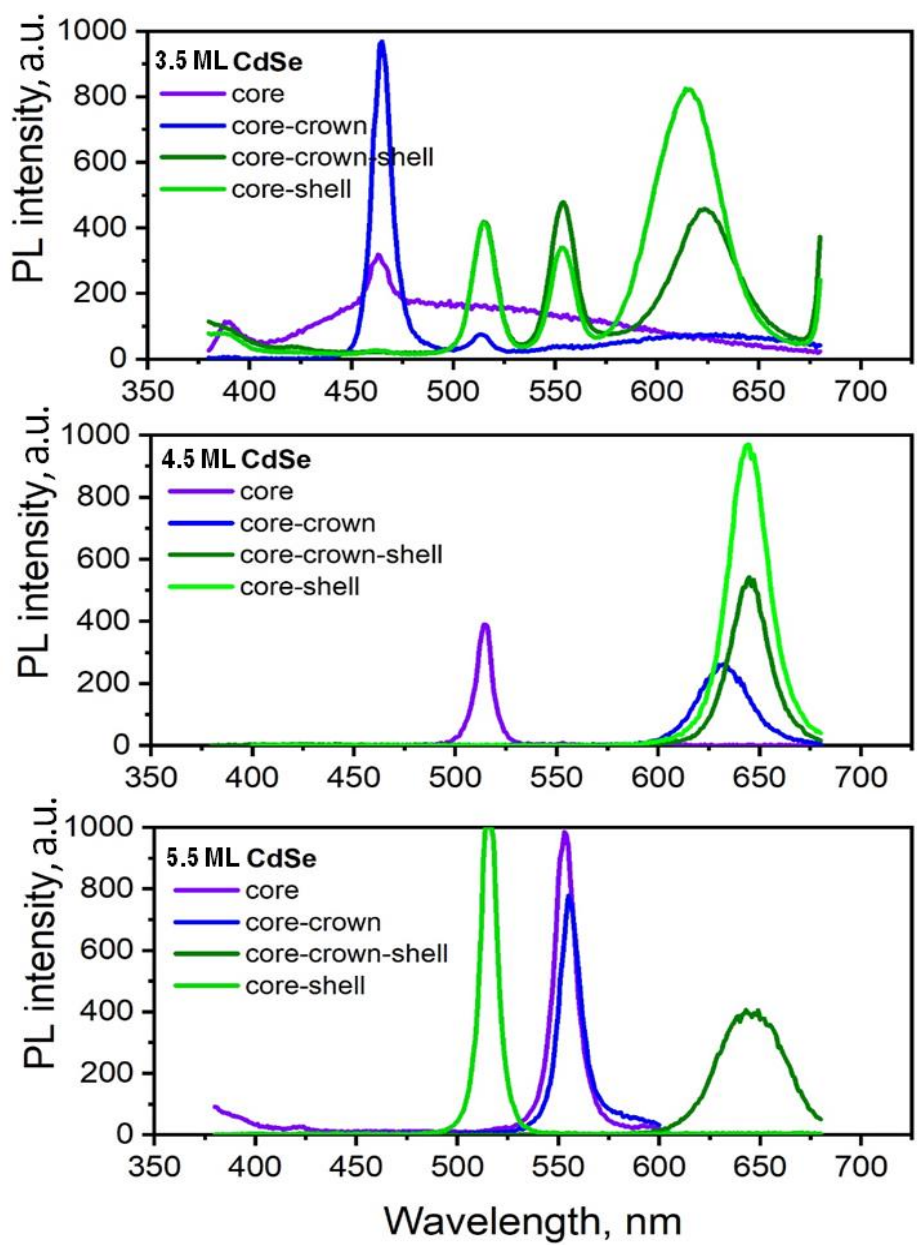

**Fig. S4.** PL emission spectra for 3.5 ML (above), 4.5 ML (middle), and 5.5 ML (below) CdSe NPLs.

## 4.4. Infrared spectroscopy

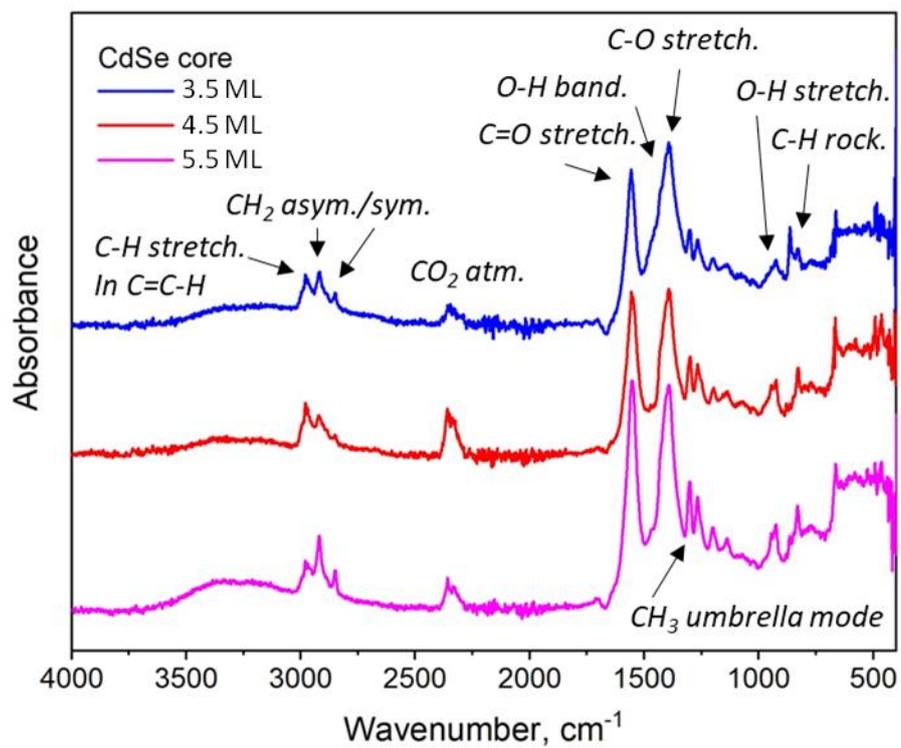

**Fig. S5.** FTIR spectra of 3.5 ML (above), 4.5 ML (middle), and 5.5 ML -5 ML CdSe core only NPLs.

## 5. Synchrotron characterization results

### 5.1. CdSe cores

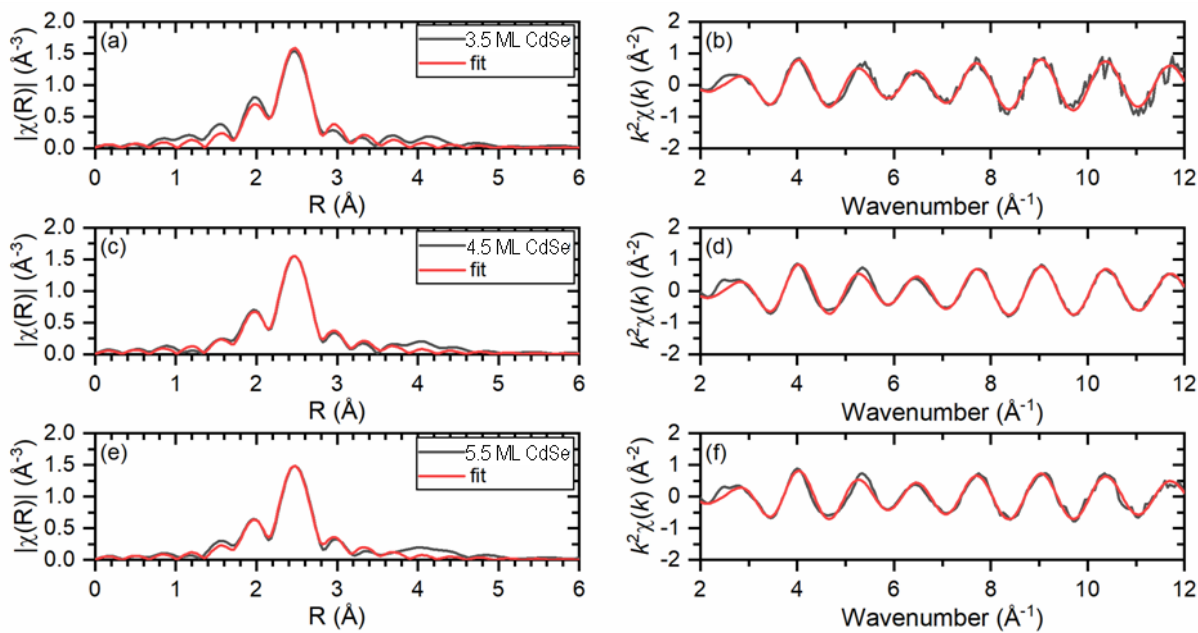

**Fig. S6.** EXAFS χ(R) function (right column) and respective Fourier transformations (left) for Se *K*-edge of 3.5 ML (a, b), 4.5 ML (c, d) and 5.5 ML (e, f) CdSe samples. The black curve represents the experimental data while red one is the best fit.

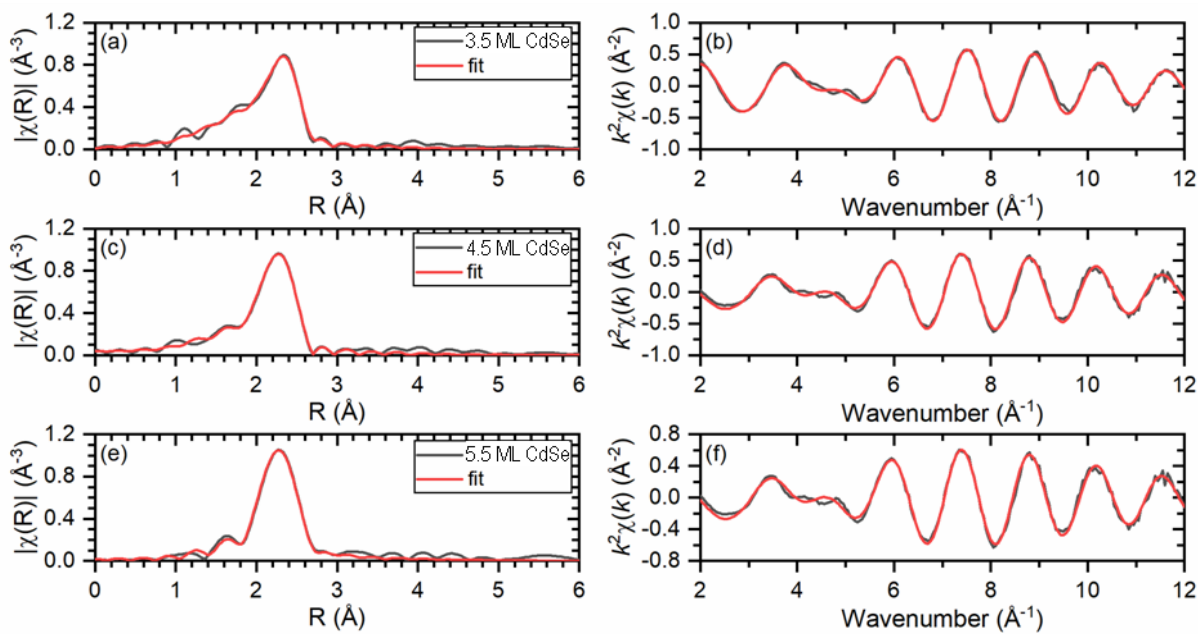

**Fig. S7.** EXAFS χ(R) function (right column) and respective Fourier transformations (left) for Cd *K*-edge of 3.5 ML (a, b), 4.5 ML (c, d) and 5.5 ML (e, f) CdSe samples. The black curve represents the experimental data while red one is the best fit.

**Table S2.** Parameters extracted from the standard 1st and 2nd shell Fourier analyses of Cd K-edge.

| NPLs | Bond | R (Å) | R err. (Å) | N | N err. | σ² | σ² err. |
|---|---|---|---|---|---|---|---|
| 3.5 ML CdSe/Cd$_x$Zn$_{1-x}$S | Cd-Se | 2.6204 | 0.0063 | 3.165 | 0.335 | 0.0068 | 0.0007 |
|  | Cd-O | 2.2480 | 0.0402 | 2.252 | 0.731 | 0.0183 | 0.0075 |
| 4.5 ML CdSe/Cd$_x$Zn$_{1-x}$S | Cd-Se | 2.6218 | 0.0047 | 3.416 | 0.266 | 0.0066 | 0.0005 |
|  | Cd-O | 2.2305 | 0.0531 | 2.172 | 0.979 | 0.0237 | 0.0105 |
| 5.5 ML CdSe/Cd$_x$Zn$_{1-x}$S | Cd-Se | 2.6273 | 0.0044 | 3.303 | 0.269 | 0.0062 | 0.0005 |
|  | Cd-O | 2.1990 | 0.0456 | 1.341 | 0.531 | 0.0150 | 0.0081 |

## 5.2 CdSe/CdS/CdZnS and CdSe/CdZnS NPLs

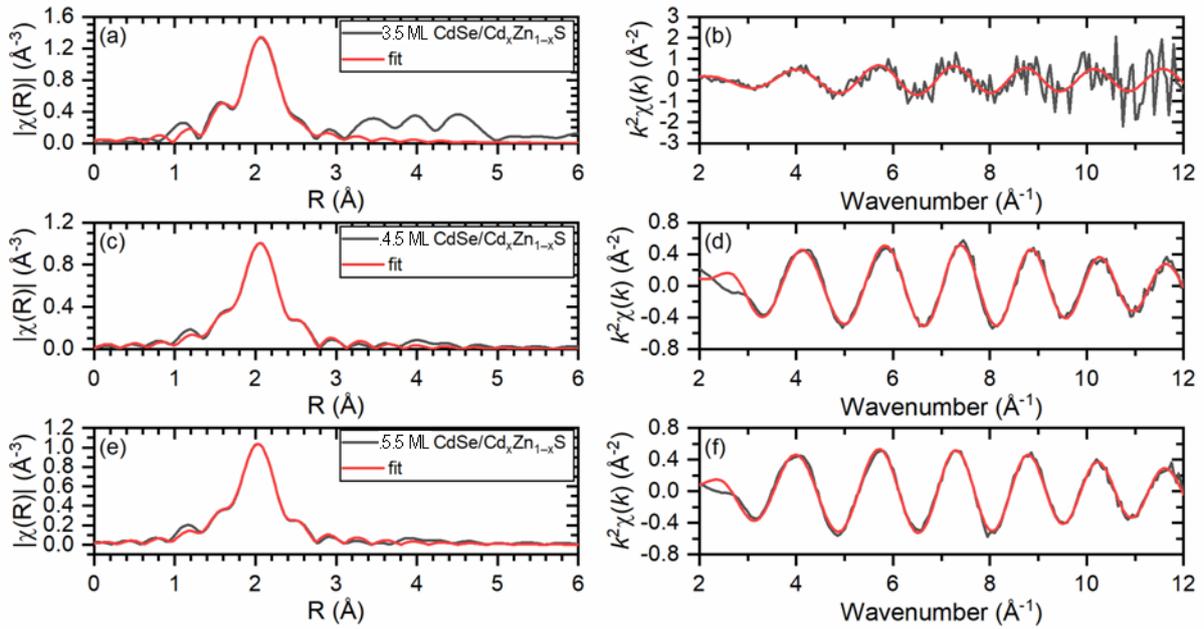

**Fig. S8.** EXAFS χ(R) function (right column) and respective Fourier transformations (left) for Cd *K*-edge of 3.5 ML (a, b), 4.5 ML (c, d) and 5.5 ML (e, f) CdSe/Cd$_x$Zn$_{1-x}$S samples. The black curve represents experimental data while red one is best fit.

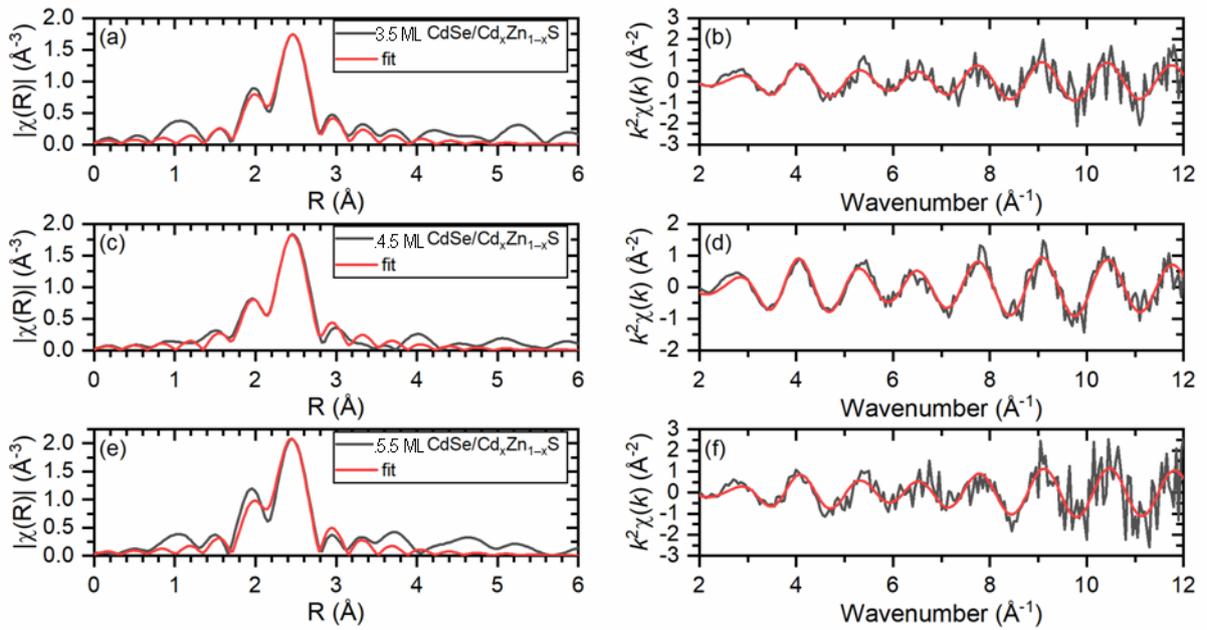

**Fig. S9.** EXAFS χ(R) function (right column) and respective Fourier transformations (left) for Se *K*-edge of 3.5 ML (a, b), 4.5 ML (c, d) and 5.5 ML (e, f) CdSe/Cd$_x$Zn$_{1-x}$S samples. The black curve represents experimental data while red one is best fit.

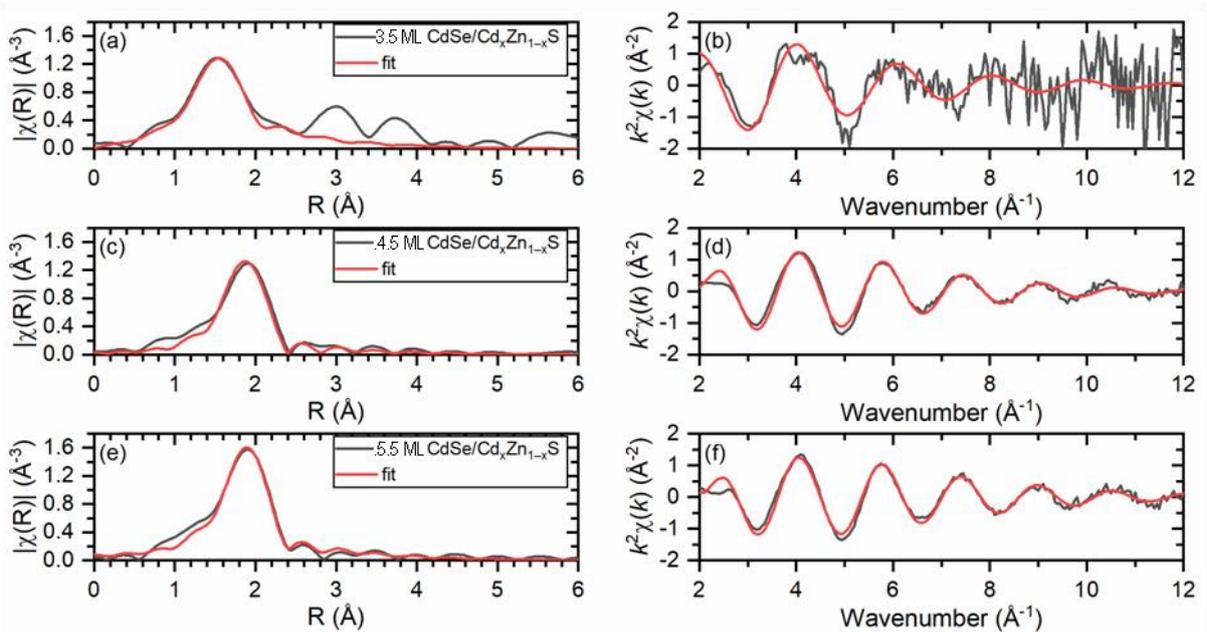

**Fig. S10.** EXAFS χ(R) function (right column) and respective Fourier transformations (left) for Zn *K*-edge of 3.5 ML (a, b), 4.5 ML (c, d) and 5.5 ML (e, f) CdSe/Cd$_x$Zn$_{1-x}$S samples. The black curve represents experimental data while red one is best fit.

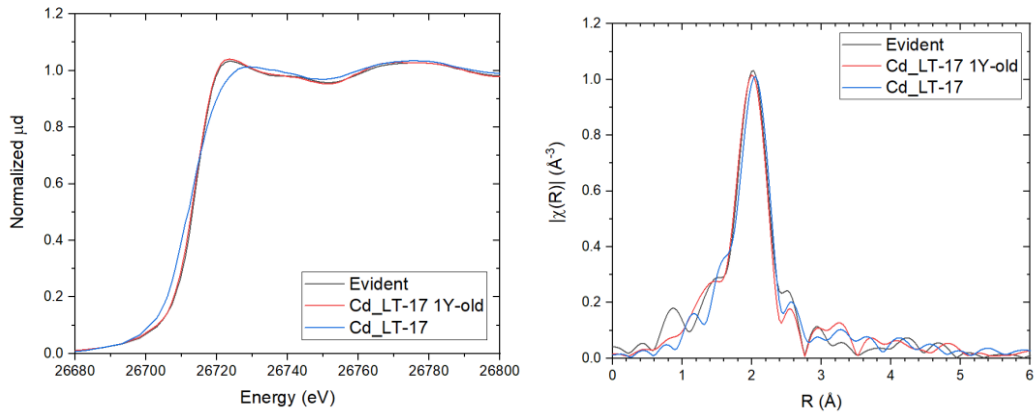

**Figure. S11.** Normalized K-edge Cd XANES spectra (*left panel*) and Fourier transformations of the Cd K-edge EXAFS plotted in radial distribution (*right panel*) of the NPLs samples after 2.5 months (blue line) and 1.5 year (red line) after the synthesis in comparison with a commercial quantum dot sample CdSe/ZnS (Evident Technologies[*]).

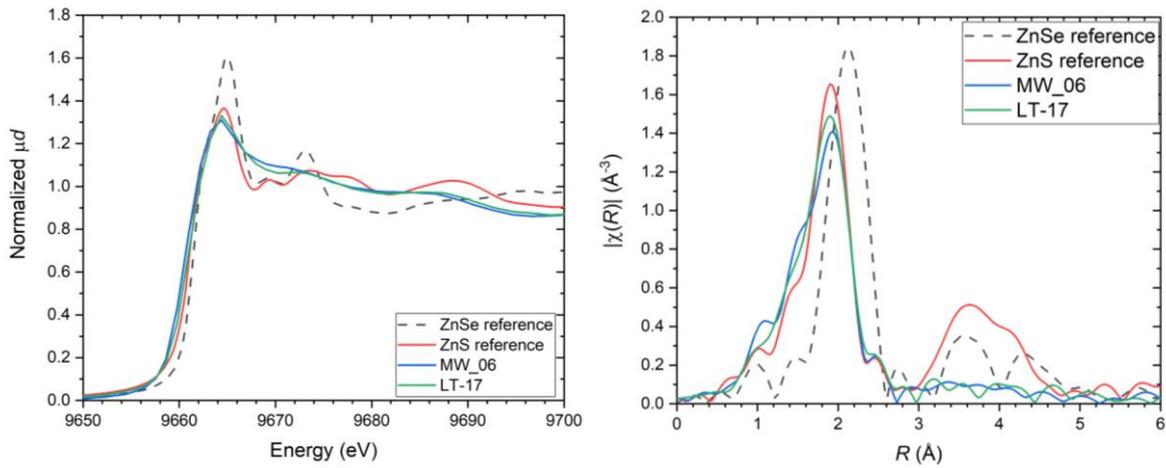

**Figure S13.** Normalized Zn *K*-edge XANES spectra (*left panel*) and respective FT of EXAFS (*right panel*) of the NPLs samples (LT-17, MW-06 and ZnS, ZnSe references).

**Table S3.** Fitting parameters* derived from the EXAFS analysis (LT-17, MW-06 & ZnS)

| Sample | R (Å) | N | $\sigma^2$ (Å$^2$) | $E_0$ (eV) |
|---|---|---|---|---|
| ZnS | 2.344 ± 0.008 | 4 | 0.0051 ± 0.0012 | 7.26 ± 0.91 |
| 4.5 ML CdSe/Cd$_x$Zn$_{1-x}$S 2.5 months | 2.334 ± 0.009 | 4.8 ± 0.4 | 0.0094 ± 0.0014 | 8.66 ± 0.80 |
| 4.5 ML CdSe/Cd$_x$Zn$_{1-x}$S 6 months | 2.339 ± 0.018 | 4.9 ± 0.8 | 0.0105 ± 0.0027 | 8.76 ± 1.53 |

**\*** see Table 1 for the designation of the reported parameters.

---

[*] Evident Technologies was a Troy, NY based nanotechnology company specializing in the commercialization of quantum dot semiconductor nanocrystals.

## 5.3 CdSe/CdS Core-Crown and CdSe/CdS/CdxZn1–xS NPLs

**Table S4**. Parameters extracted from the standard 1st shell FT analysis of Se K-edge for 3.5, 4.5 and 5.5 ML CdSe/CdS core-crown NPLs.

| Name | Bonding | R (Å) | R error (Å) | N | N error | $\sigma^2$ | $\sigma^2$ error |
|---|---|---|---|---|---|---|---|
| 3.5 ML CdSe/CdS | Se-Cd | 2.6102 | 0.0058 | 3.020 | 0.307 | 0.0050 | 0.0008 |
| 4.5 ML CdSe/CdS | Se-Cd | 2.6152 | 0.0032 | 3.147 | 0.173 | 0.0051 | 0.0004 |
| 5.5 ML CdSe/CdS | Se-Cd | 2.6171 | 0.0031 | 3.198 | 0.174 | 0.0052 | 0.0004 |

**Table S5**. Parameters extracted from the standard 1st shell FT analysis of Se K-edge for 3.5, 4.5 and 5.5 ML CdSe/CdS/Cd$_x$Zn$_{1-x}$S.

| Name | Bonding | R (Å) | R error (Å) | N | N error | $\sigma^2$ | $\sigma^2$ error |
|---|---|---|---|---|---|---|---|
| 3.5 ML CdSe/CdS/Cd$_x$Zn$_{1-x}$S | Se-Cd | 2.5605 | 0.0132 | 3.394 | 0.601 | 0.0051 | 0.0021 |
| 4.5 ML CdSe/CdS/Cd$_x$Zn$_{1-x}$S | Se-Cd | 2.6375 | 0.0206 | 3.601 | 0.935 | 0.0066 | 0.0032 |
| 5.5 ML CdSe/CdS/Cd$_x$Zn$_{1-x}$S | Se-Cd | 2.5933 | 0.0188 | 3.041 | 0.808 | 0.0045 | 0.0030 |